\documentclass{article}

\usepackage{microtype}
\usepackage{graphicx}
\usepackage{subcaption, multirow}
\usepackage{booktabs} % for professional tables
\usepackage[table]{xcolor}

\usepackage[hyphens]{url}
\usepackage{hyperref}

\usepackage{amsmath}
\usepackage{amssymb}
\usepackage{amsthm}
\usepackage{paralist}
\usepackage{enumerate}

\usepackage{diagbox}
\usepackage{array}
\newcolumntype{L}[1]{>{\raggedright\let\newline\\\arraybackslash\hspace{0pt}}m{#1}}
\newcolumntype{C}[1]{>{\centering\let\newline\\\arraybackslash\hspace{0pt}}m{#1}}
\newcolumntype{R}[1]{>{\raggedleft\let\newline\\\arraybackslash\hspace{0pt}}m{#1}}
\usepackage{xcolor}

\newcommand{\STAB}[1]{\begin{tabular}{@{}c@{}}#1\end{tabular}}
\usepackage[accepted]{icml2020}

\icmltitlerunning{Two Routes to Scalable Credit Assignment without Weight Symmetry}

\begin{document}

\twocolumn[

\icmltitle{Two Routes to Scalable Credit Assignment without Weight Symmetry}



\icmlsetsymbol{equal}{*}

\begin{icmlauthorlist}
\icmlauthor{Daniel Kunin}{equal,icme}
\icmlauthor{Aran Nayebi}{equal,neurodept}
\icmlauthor{Javier Sagastuy-Brena}{equal,icme}
\\
\icmlauthor{Surya Ganguli}{apdept}
\icmlauthor{Jonathan M. Bloom}{broad,cell}
\icmlauthor{Daniel L. K. Yamins}{psychdept,csdept,wtsni}
\end{icmlauthorlist}

% \icmlaffiliation{snail}{Stanford Neuroscience and Artificial Intelligence Laboratory, Stanford University}
% \icmlaffiliation{ndcl}{Neural Dynamics and Computation Laboratory, Stanford University}
\icmlaffiliation{icme}{Institute for Computational and Mathematical Engineering, Stanford University}
\icmlaffiliation{neurodept}{Neurosciences PhD Program, Stanford University}
\icmlaffiliation{apdept}{Department of Applied Physics, Stanford University}
\icmlaffiliation{broad}{Broad Institute of MIT and Harvard, Cambridge, MA}
\icmlaffiliation{cell}{Cellarity, Cambridge, MA}
\icmlaffiliation{psychdept}{Department of Psychology, Stanford University}
\icmlaffiliation{csdept}{Department of Computer Science, Stanford University}
\icmlaffiliation{wtsni}{Wu Tsai Neurosciences Institute, Stanford University}

\icmlcorrespondingauthor{Aran Nayebi}{anayebi@stanford.edu}
\icmlcorrespondingauthor{Daniel Kunin}{kunin@stanford.edu}
\icmlcorrespondingauthor{Javier Sagastuy-Brena}{jvrsgsty@stanford.edu}

\icmlkeywords{}

\vskip 0.3in
]



\printAffiliationsAndNotice{\icmlEqualContribution} % otherwise use the standard text.

\begin{abstract}

The neural plausibility of backpropagation has long been disputed, primarily for its use of non-local weight transport --- the biologically dubious requirement that one neuron instantaneously measure the synaptic weights of another. 
Until recently, attempts to create local learning rules that avoid weight transport have typically failed in the large-scale learning scenarios where backpropagation shines, e.g. ImageNet categorization with deep convolutional networks.
Here, we investigate a recently proposed local learning rule that yields competitive performance with backpropagation and find that it is highly sensitive to metaparameter choices, requiring laborious tuning that does not transfer across network architecture.
Our analysis indicates the underlying mathematical reason for this instability, allowing us to identify a more robust local learning rule that better transfers without metaparameter tuning.
Nonetheless, we find a performance and stability gap between this local rule and backpropagation that widens with increasing model depth.
We then investigate several non-local learning rules that relax the need for instantaneous weight transport into a more biologically-plausible ``weight estimation'' process, showing that these rules match state-of-the-art performance on deep networks and operate effectively in the presence of noisy updates.
Taken together, our results suggest two routes towards the discovery of neural implementations for credit assignment without weight symmetry: further improvement of local rules so that they perform consistently across architectures and the identification of biological implementations for non-local learning mechanisms.
\end{abstract}

\section{Introduction}
\label{intro}

Backpropagation is the workhorse of modern deep learning and the only known learning algorithm that allows multi-layer networks to train on large-scale tasks.
However, any exact implementation of backpropagation is inherently non-local, requiring instantaneous weight transport in which backward error-propagating weights are the transpose of the forward inference weights. 
This violation of locality is biologically suspect because there are no known neural mechanisms for instantaneously coupling distant synaptic weights. 
Recent approaches such as feedback alignment \cite{lillicrap_random_2016} and weight mirror \cite{akrout_deep_2019} have identified circuit mechanisms that seek to approximate backpropagation while circumventing the weight transport problem. 
However, these mechanisms either fail to operate at large-scale \cite{bartunov_assessing_2018} or, as we demonstrate, require complex and fragile metaparameter scheduling during learning. 
Here we present a unifying framework spanning a space of learning rules that allows for the systematic identification of robust and scalable alternatives to backpropagation. 

To motivate these rules, we replace tied weights in backpropagation with a regularization loss on untied forward and backward weights. 
The forward weights parametrize the global cost function, the backward weights specify a descent direction, and the regularization constrains the relationship between forward and backward weights. 
As the system iterates, forward and backward weights dynamically align, giving rise to a pseudogradient. 
Different regularization terms are possible within this framework. 
Critically, these regularization terms decompose into geometrically natural primitives, which can be parametrically recombined to construct a diverse space of credit assignment strategies. 
This space encompasses existing approaches (including feedback alignment and weight mirror), but also elucidates novel learning rules. 
We show that several of these new strategies are competitive with backpropagation on real-world tasks (unlike feedback alignment), without the need for complex metaparameter tuning (unlike weight mirror). 
These learning rules can thus be easily deployed across a variety of neural architectures and tasks. 
Our results demonstrate how high-dimensional error-driven learning can be robustly performed in a biologically motivated manner.

\section{Related Work}
\label{sec:related}

Soon after \citet{rumelhart_learning_1986} published the backpropagation algorithm for training neural networks, its plausibility as a learning mechanism in the brain was contended \cite{crick_recent_1989}.  The main criticism was that backpropagation requires exact transposes to propagate errors through the network and there is no known physical mechanism for such an ``operation" in the brain. 
This is known as the \textit{weight transport problem} \cite{grossberg_competitive_1987}. Since then many credit assignment strategies have proposed circumventing the problem by introducing a distinct set of feedback weights to propagate the error backwards. Broadly speaking, these proposals fall into two groups: those that encourage symmetry between the forward and backward weights \cite{lillicrap_random_2016,nokland_direct_2016,bartunov_assessing_2018,liao_how_2016,xiao_biologically-plausible_2019,moskovitz_feedback_2018,akrout_deep_2019}, and those that encourage preservation of information between neighboring network layers \cite{bengio_how_2014, lee_difference_2015, bartunov_assessing_2018}.

The latter approach, sometimes referred to as target propagation, encourages the backward weights to locally invert the forward computation \cite{bengio_how_2014}.  
Variants of this approach such as difference target propagation \cite{lee_difference_2015} and simplified difference target propagation \cite{bartunov_assessing_2018} differ in how they define this inversion property.  
While some of these strategies perform well on shallow networks trained on MNIST and CIFAR10, they fail to scale to deep networks trained on ImageNet \cite{bartunov_assessing_2018}.

A different class of credit assignment strategies focuses on encouraging or enforcing symmetry between the weights, rather than preserving information.
\citet{lillicrap_random_2016} introduced a strategy known as feedback alignment in which backward weights are chosen to be fixed random matrices. 
Empirically, during learning, the forward weights partially align themselves to their backward counterparts, so that the latter transmit a meaningful error signal. \citet{nokland_direct_2016} introduced a variant of feedback alignment where the error signal could be transmitted across long range connections. However, for deeper networks and more complex tasks, the performance of feedback alignment and its variants break down \cite{bartunov_assessing_2018}.

\citet{liao_how_2016} and \citet{xiao_biologically-plausible_2019} took an alternative route to relaxing the requirement for exact weight symmetry by transporting just the sign of the forward weights during learning. \citet{moskovitz_feedback_2018} combined sign-symmetry and feedback alignment with additional normalization mechanisms. These methods outperform feedback alignment on scalable tasks, but still perform far worse than backpropation. It is also not clear that instantaneous sign transport is more biologically plausible than instantaneous weight transport.

More recently, \citet{akrout_deep_2019} introduced weight mirror (WM), a learning rule that incorporates dynamics on the backward weights to improve alignment throughout the course of training. Unlike previous methods, weight mirror achieves backpropagation level performance on ResNet-18 and ResNet-50 trained on ImageNet.

Concurrently, \citet{kunin_loss_2019} suggested training the forward and backward weights in each layer as an encoder-decoder pair, based on their proof that $L_2$-regularization induces symmetric weights for linear autoencoders.  This approach incorporates ideas from both information preservation and weight symmetry. 

A complementary line of research \cite{Xie2003Equivalence, Scellier2017Equilibrium, bengio2017STDP, guerguiev_towards_2017, whittington2017approximation,  Sacramento2018Dendritic, guerguiev_spike-based_2019} investigates how learning rules, even those that involve weight transport, could be implemented in a biologically mechanistic manner, such as using  spike-timing dependent plasticity rules and obviating the need for distinct phases of training. 
In particular, \citet{guerguiev_spike-based_2019} show that key steps in the \citet{kunin_loss_2019} regularization approach could be implemented by a spike-based mechanism for approximate weight transport.

In this work, we extend this regularization approach to formulate a more general framework of credit assignment strategies without weight symmetry, one that encompasses existing and novel learning rules.
Our core result is that the best of these strategies are substantially more robust across architectures and metaparameters than previous proposals. 

\section{Regularization Inspired Learning Rule Framework}
\label{sec:framework}

\begin{figure}[tb]
\begin{center}
\centerline{
\includegraphics[width=0.75\linewidth]{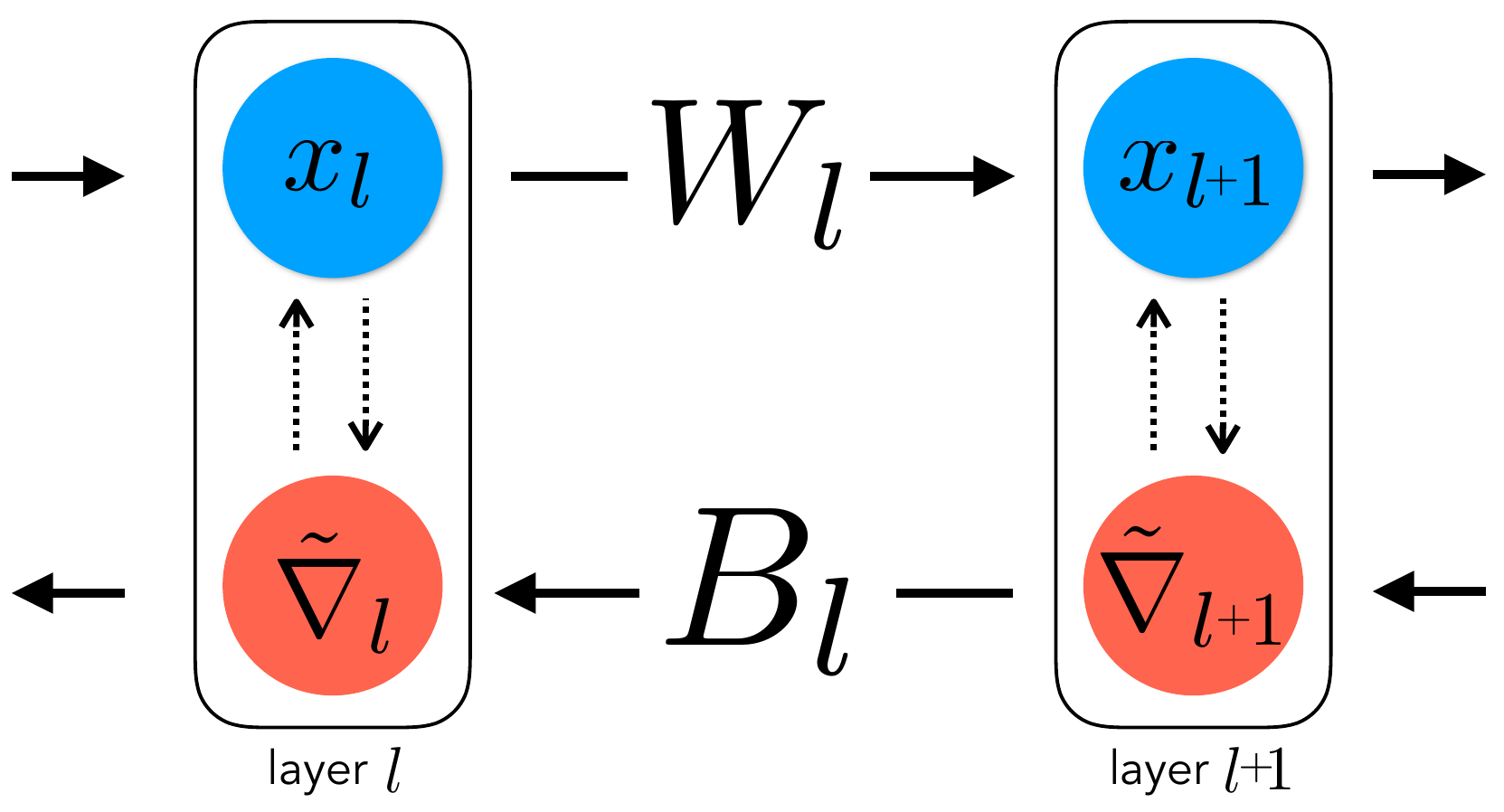}
}
\caption{\textbf{Notational diagram.} The forward weight $W_l$ propagates the input signal $x_l$ downstream through the network.  The backward weight $B_l$ propagates the pseudogradient $\widetilde{\nabla}_{l+1}$ of $\mathcal{J}$ upstream through the network.  The regularization function $\mathcal{R}$ is constructed layer-wise from $x_l$, $x_{l+1}$, $W_l$ and $B_l$. Similar to \citet{akrout_deep_2019}, we assume lateral pathways granting the backward weights access to $x$ and the forward weights access to $\widetilde{\nabla}$.  Biases and non-linearities are omitted from the diagram.}
\label{fig:conceptual-framework}
\end{center}
\vskip -0.3in
\end{figure}

We consider the credit assignment problem for neural networks as a layer-wise regularization problem. 
We consider a network parameterized by forward weights $\theta_f$ and backward weights $\theta_b$. 
Informally, the network is trained on the sum of a global task function $\mathcal{J}$ and a layer-wise regularization function\footnote{ $\mathcal{R}$ is not regularization in the traditional sense, as it does not directly penalize the forward weights $\theta_f$ from the cost function $\mathcal{J}$.} $\mathcal{R}$:
\begin{equation*}
\mathcal{L}(\theta_f,\theta_b) = \mathcal{J}(\theta_f) + \mathcal{R}(\theta_b).
\end{equation*}
Formally, every step of training consists of two updates, one for the forward weights and one for the backward weights. 
The forward weights are updated according to the error signal on $\mathcal{J}$ propagated through the network by the backward weights, as illustrated in Fig.~\ref{fig:conceptual-framework}. 
The backward weights are updated according to gradient descent on $\mathcal{R}$. 
$$\Delta \theta_f \propto \widetilde{\nabla}J \qquad \Delta \theta_b \propto \nabla R$$
Thus, $\mathcal{R}$ is responsible for introducing dynamics on the backward weights, which in turn impacts the dynamics of the forward weights. 
The functional form of $\mathcal{R}$ gives rise to different learning rules and in particular the locality of a given learning rule depends solely on the locality of the computations involved in $\mathcal{R}$.

\subsection{Regularization Primitives}
\label{sec:framework-primitives}

In this work, the regularization function $\mathcal{R}$ is built from a set of simple \emph{primitives} $\mathcal{P}$, which at any given layer $l$ are functions of the forward weight $W_l \in \theta_f$, backward weight $B_l \in \theta_b$, layer input $x_l$, and layer output $x_{l+1}$ as depicted in Fig.~\ref{fig:conceptual-framework}.  These primitives are biologically motivated components with strong geometric interpretations, from which more complex learning rules may be algebraically constructed.  

The primitives we use, displayed in Table~\ref{tab:prim}, can be organized into two groups: those that involve purely local operations and those that involve at least one non-local operation.
To classify the primitives, we use the criteria for locality described in \citet{whittington2017approximation}: 
(1) \textit{Local computation}. Computations only involve synaptic weights acting on their associated inputs.
(2) \textit{Local plasticity}.  Weight modifications only depend on pre-synaptic and post-synaptic activity.
A primitive is local if it satisfies both of these constraints and non-local otherwise.  

We introduce three \textbf{local primitives}: $\mathcal{P}^{\text{decay}}$, $\mathcal{P}^{\text{amp}}$, and $\mathcal{P}^{\text{null}}$. 
The \textit{decay} primitive can be understood as a form of energy efficiency penalizing the Euclidean norm of the backward weights.  The \textit{amp} primitive promotes alignment of the layer input $x_l$ with the reconstruction $B_lx_{l+1}$. The \textit{null} primitive imposes sparsity in the layer-wise activity through a Euclidean norm penalty on the reconstruction $B_lx_{l+1}$.

We consider two \textbf{non-local primitives}: $\mathcal{P}^{\text{sparse}}$ and $\mathcal{P}^{\text{self}}$. 
The \textit{sparse} primitive promotes energy efficiency by penalizing the Euclidean norm of the activation $x_l^\intercal B_l$.  This primitive fails to meet the \textit{local computation} constraint, as $B_l$ describes the synaptic connections from the $l+1$ layer to the $l$ layer and therefore cannot operate on the input $x_l$. 

The \textit{self} primitive promotes alignment of the forward and backward weights by directly promoting their inner product.  This primitive fails the \textit{local plasticity} constraint, as its gradient necessitates that the backward weights can exactly measure the strengths of their forward counterparts.  

\begin{table}
\centering
\begin{tabular}{cccc} \toprule
    & \multicolumn{1}{c}{Local} & \multicolumn{1}{c}{$\mathcal{P}_l$} & $\nabla\mathcal{P}_l$\\ \midrule 
    & decay & $\frac{1}{2}||B_l||^2$ & $B_l$\\
    & amp & $-\mathrm{tr}(x_l^\intercal B_l x_{l+1})$ & $-x_lx_{l+1}^\intercal$ \\
    & null & $\frac{1}{2}||B_lx_{l+1}||^2$ & $B_lx_{l+1}x_{l+1}^\intercal$\\
    \midrule
    & \multicolumn{1}{c}{Non-local} & \multicolumn{1}{c}{$\mathcal{P}_l$} & $\nabla\mathcal{P}_l$\\ \midrule
    & sparse & $\frac{1}{2}||x_l^\intercal B_l||^2$ & $x_lx_l^\intercal B_l$\\
    & self & $-\mathrm{tr}(B_l W_l)$ & $-W_l^\intercal$ \\\bottomrule
\end{tabular}
%%%%%%%%%%%
\caption{\textbf{Regularization primitives.} Mathematical expressions for local and non-local primitives and their gradients with respect to the backward weight $B_l$.
Note, both $x_l$ and $x_{l+1}$ are the post-nonlinearity rates of their respective layers.
\label{tab:prim}}
\vskip -0.1in
\end{table}

\subsection{Building Learning Rules from Primitives}
\label{sec:framework-learning_circuits}

These simple primitives can be linearly combined to encompass existing credit assignment strategies, while also elucidating natural new approaches. 

\textbf{Feedback alignment (FA)} \cite{lillicrap_random_2016} corresponds to no regularization, $\mathcal{R}_{\text{FA}} \equiv 0$, effectively fixing the backward weights at their initial random values\footnote{We explore the consequences of this interpretation analytically in Appendix~\ref{sec:analysis-beyond_fa}.}.

The \textbf{weight mirror (WM)} \cite{akrout_deep_2019} update, $\Delta B_l = \eta x_lx_{l+1}^\intercal - \lambda_{\text{WM}} B_l$, where $\eta$ is the learning rate and $\lambda_{\text{WM}}$ is a weight decay constant, corresponds to gradient descent on the layer-wise regularization function
$$\mathcal{R}_{\text{WM}} = \sum_{l \in \text{layers}} \alpha\mathcal{P}^{\text{amp}}_l + \beta\mathcal{P}^{\text{decay}}_l,$$
for $\alpha = 1$ and $\beta = \frac{\lambda_{\text{WM}}}{\eta}$.

If we consider primitives that are functions of the pseudogradients $\widetilde{\nabla}_{l}$ and $\widetilde{\nabla}_{l+1}$, then the \textbf{Kolen-Pollack (KP)} algorithm, originally proposed by \citet{Kolen1994backpropagation} and modified by \citet{akrout_deep_2019}, can be understood in this framework as well.
See Appendix~\ref{sec:kolen-pollack} for more details.

The range of primitives also allows for learning rules not yet investigated in the literature. 
In this work, we introduce several such novel learning rules, including Information Alignment (IA), Symmetric Alignment (SA), and Activation Alignment (AA).  
Each of these strategies is defined by a layer-wise regularization function composed from a linear combination of the primitives (Table~\ref{tab:taxonomy}).  
Information Alignment is a purely local rule, but unlike feedback alignment or weight mirror, contains the additional null primitive. 
In \S\ref{sec:local-learning-rules}, we motivate this addition theoretically, and show empirically that it helps make IA a higher-performing and substantially more stable learning rule than previous local strategies.
SA and AA are both non-local, but as shown in \S\ref{sec:non-local-learning-rules} perform even more robustly than any local strategy we or others have found, and may be implementable by a type of plausible biological mechanism we call ``weight estimation.''

\begin{table}[tb]
\setlength\tabcolsep{4pt}
\hspace*{-0.25cm}
\centering
\begin{tabular}{@{}|c|l|c|c|c|c|c|@{}} \toprule
    & \multicolumn{1}{c|}{\textbf{Alignment}} & \multicolumn{1}{c|}{$\mathcal{P}^{\text{decay}}$} & \multicolumn{1}{c|}{$\mathcal{P}^{\text{amp}}$} & \multicolumn{1}{c|}{$\mathcal{P}^{\text{null}}$} & \multicolumn{1}{c|}{$\mathcal{P}^{\text{sparse}}$} & \multicolumn{1}{c|}{$\mathcal{P}^{\text{self}}$} \\ \hline
    \multirow{3}{*}{\STAB{\rotatebox[origin=c]{90}{\small Local}}}
    & Feedback & & & & & \\
    & Weight Mirror & \checkmark &  \checkmark & &  & \\
    & Information & \checkmark & \checkmark & \checkmark & & \\\midrule
    \multirow{3}{*}{\STAB{\rotatebox[origin=c]{90}{\small \shortstack{Non-\\Local}\hspace{-0.8em}}}}
    & Symmetric & \checkmark & & & & \checkmark \\
    & Activation & & \checkmark & & \checkmark & \\
    \bottomrule
\end{tabular}
\caption{\textbf{Taxonomy of learning rules} based on the locality and composition of their primitives. \label{tab:taxonomy}}
\vskip -0.2in
\end{table}

\subsection{Evaluating Learning Rules}
\label{sec:guidelines}
For all the learning rules, we evaluate two desirable target metrics.

\textbf{Task Performance.} 
Performance-optimized CNNs on ImageNet provide the most effective quantitative description of neural responses of cortical neurons throughout the primate ventral visual pathway~\cite{yamins2014performance, cadena2019deep}, indicating the biological relevance of task performance. 
Therefore, our first desired target will be ImageNet top-1 validation accuracy, in line with \citet{bartunov_assessing_2018}.

\textbf{Metaparameter Robustness.}
Extending the proposal of \citet{bartunov_assessing_2018}, we also consider whether a proposed learning rule's metaparameters, such as learning rate and batch size, transfer across architectures. 
Specifically, when we optimize for metaparameters on a given architecture (e.g. ResNet-18), we will fix these metaparameters and use them to train both deeper (e.g. ResNet-50) and different variants (e.g. ResNet-v2).
Therefore, our second desired target will be ImageNet top-1 validation accuracy \emph{across} models for \emph{fixed} metaparameters.

\section{Local Learning Rules}
\label{sec:local-learning-rules}

\textbf{Instability of Weight Mirror.} \citet{akrout_deep_2019} report that the weight mirror update rule matches the performance of backpropagation on ImageNet categorization.
The procedure described in \citet{akrout_deep_2019} involves not just the weight mirror rule, but a number of important additional training details, including alternating learning modes and using layer-wise Gaussian input noise. 
After reimplementing this procedure in detail, and using their prescribed metaparameters for the ResNet-18 architecture, the best top-1 validation accuracy we were able to obtain was 63.5\% ($\mathcal{R}_{\text{WM}}$ in Table~\ref{tab:hp-local}), substantially below the reported performance of 69.73\%.
To try to account for this discrepancy, we considered the possibility that the metaparameters were incorrectly set.
We thus performed a large-scale metaparameter search over the continuous $\alpha$, $\beta$, and the standard deviation $\sigma$ of the Gaussian input noise, jointly optimizing these parameters for ImageNet validation set performance using a Bayesian Tree-structured Parzen Estimator (TPE) \cite{bergstra_tpe_2011}. 
After considering 824 distinct settings (see Appendix~\ref{sup:hp-ss-details} for further details), the optimal setting achieved a top-1 performance of 64.07\% ($\mathcal{R}^{\text{TPE}}_{\text{WM}}$ in Table~\ref{tab:hp-local}), still substantially below the reported performance in \citet{akrout_deep_2019}.

\begin{figure}[tb]
\centering
\includegraphics[width=1.0\columnwidth]{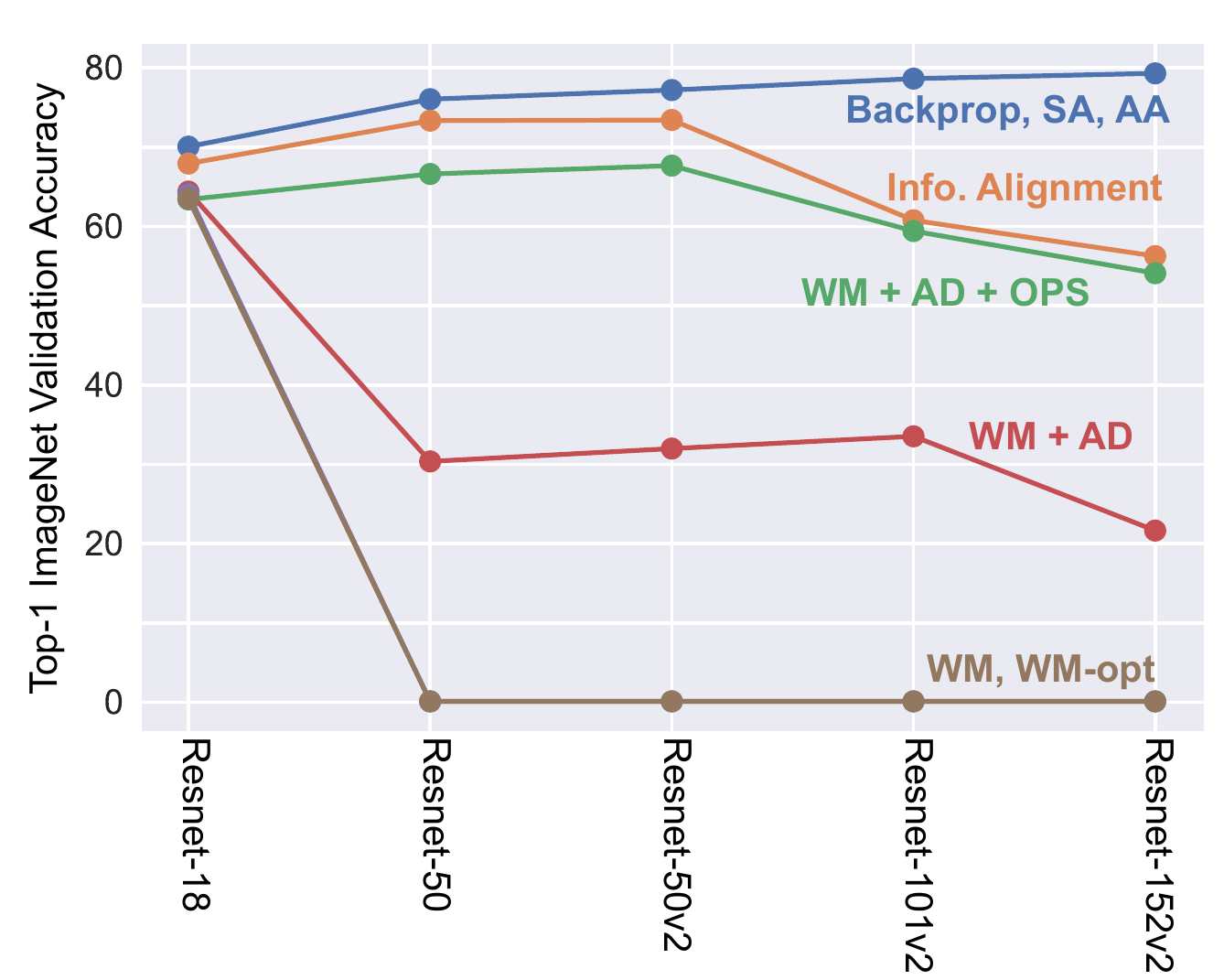}
\caption{\textbf{Performance of local and non-local rules across architectures.} We fixed the categorical and continuous metaparameters for ResNet-18 and applied them directly to deeper and different ResNet variants (e.g. v2). A performance of 0.001 indicates the alignment loss became NaN within the first thousand steps of training. Our local rule Information Alignment (IA) consistently outperforms the other \emph{local} alternatives across architectures, despite not being optimized for these architectures. The non-local rules, Symmetric Alignment (SA) and Activation Alignment (AA), consistently perform as well as backpropagation.}
\label{fig:hp-deeper}
\vskip -0.1in
\end{figure}

Considering the second metric of robustness, we found that the WM learning rule is very sensitive to metaparameter tuning. 
Specifically, when using either the metaparameters prescribed for ResNet-18 in \citet{akrout_deep_2019} or those from our metaparameter search, directly attempting to train other network architectures failed entirely (Fig.~\ref{fig:hp-deeper}, brown line).

Why is weight mirror under-performing backpropagation on both performance and robustness metrics?
Intuition can be gained by simply considering the functional form of $\mathcal{R}_{\text{WM}}$, which can become \emph{arbitrarily} negative even for fixed values of the forward weights.
$\mathcal{R}_{\text{WM}}$ is a combination of a primitive which depends on the input ($\mathcal{P}^{\text{amp}}$) and a primitive which is independent of the input ($\mathcal{P}^{\text{decay}}$).
Because of this, the primitives of weight mirror and their gradients may operate at different scales and careful metaparameter tuning must be done to balance their effects.
This instability can be made precise by considering the dynamical system given by the symmetrized gradient flow on $\mathcal{R}_{\text{WM}}$ at a given layer $l$.

\begin{figure}[tb]
\begin{subfigure}{0.32\columnwidth}
    \centering
    \includegraphics[width=\textwidth]{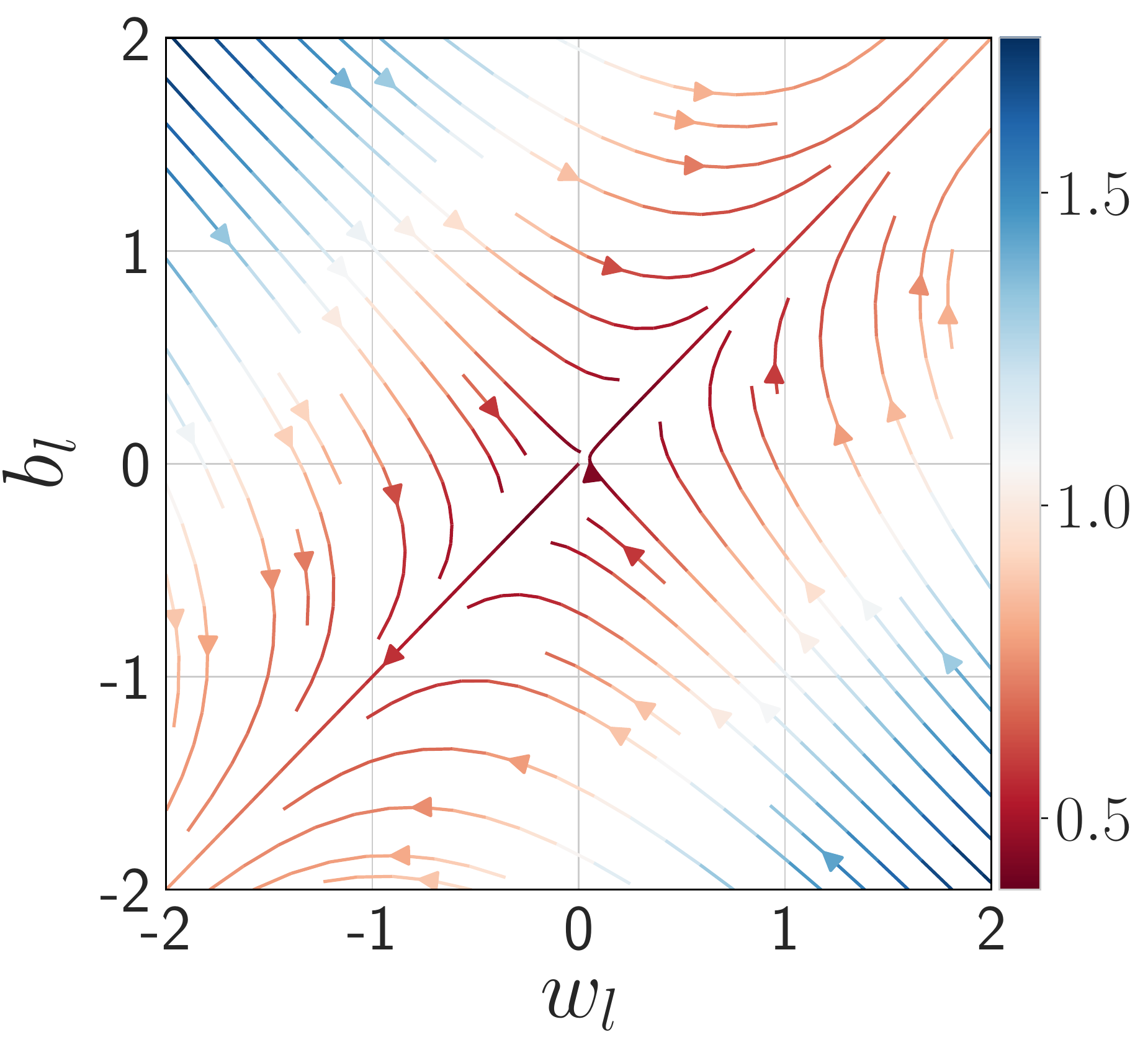}
    \caption{$\lambda_{\text{WM}} < x_l^2$}
\end{subfigure}
\begin{subfigure}{0.32\columnwidth}
    \centering
    \includegraphics[width=\textwidth]{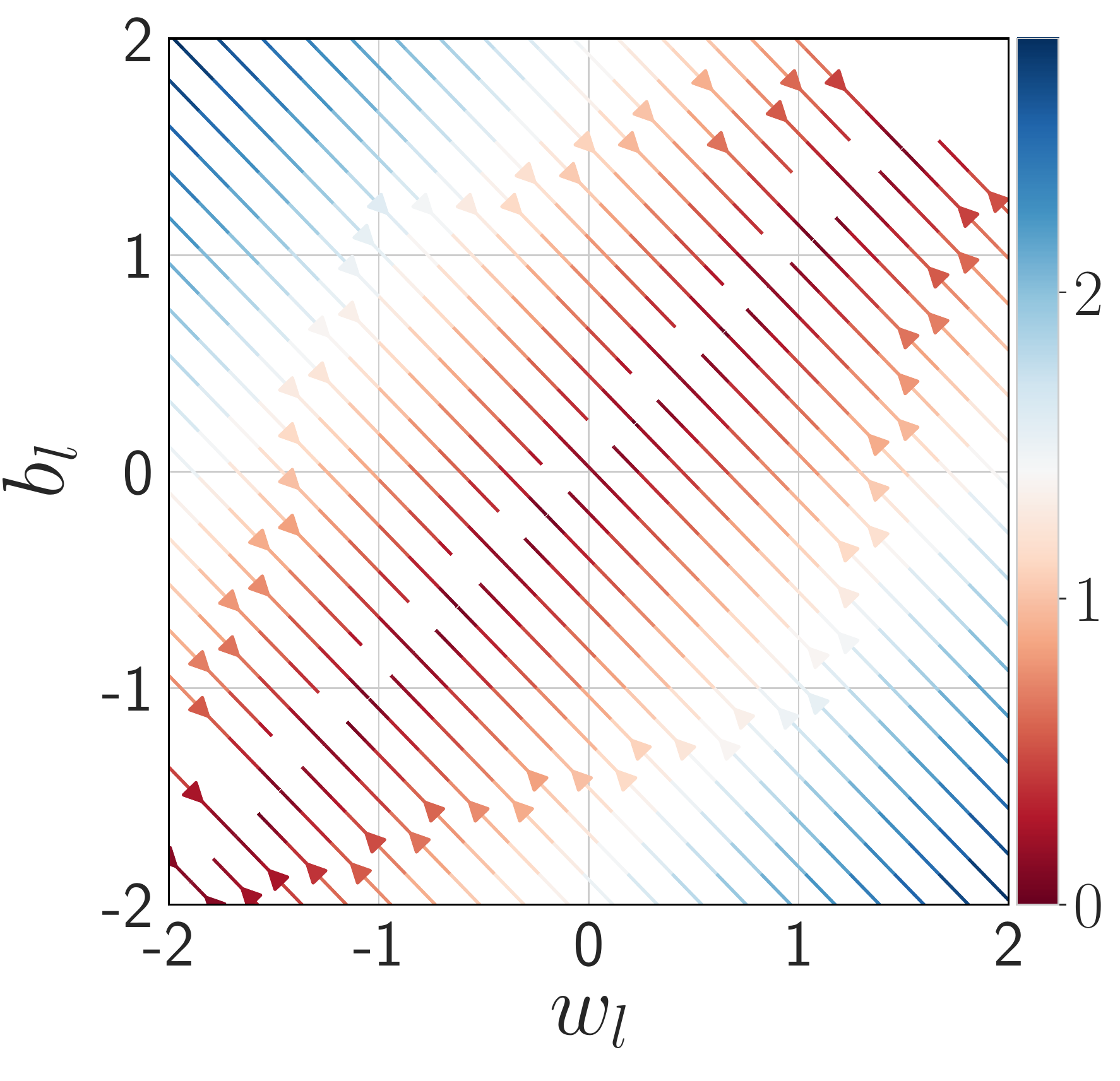}
    \caption{$\lambda_{\text{WM}} = x_l^2$}
\end{subfigure}
\begin{subfigure}{0.32\columnwidth}
    \centering
    \includegraphics[width=\textwidth]{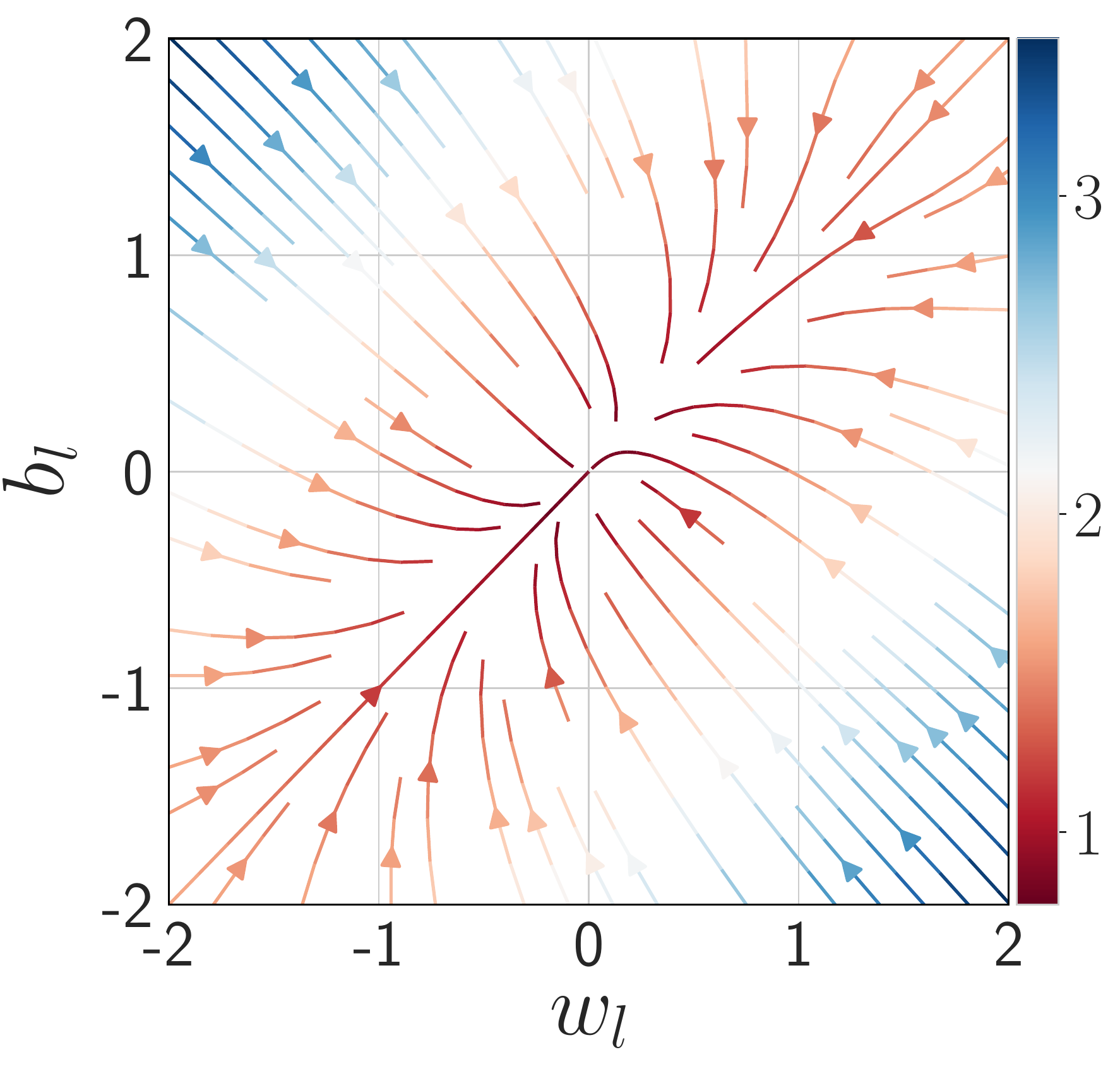}
    \caption{$\lambda_{\text{WM}} > x_l^2$}
\end{subfigure}
\vskip 0.1in
\begin{subfigure}{0.32\columnwidth}
    \centering
    \includegraphics[width=\textwidth]{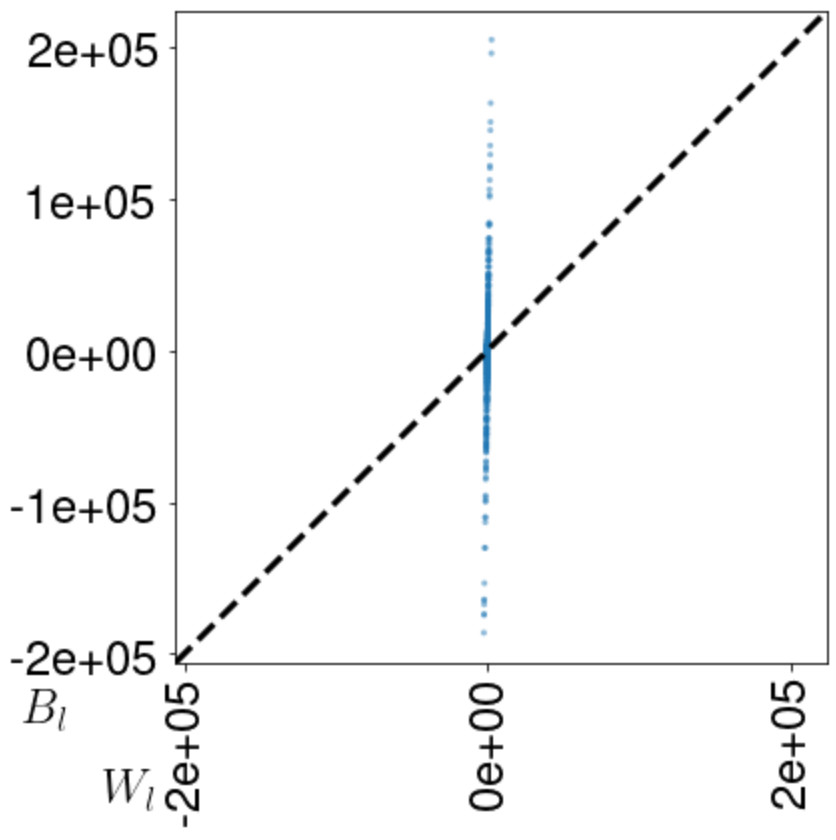}
    \caption{Conv 1}
\end{subfigure}
\begin{subfigure}{0.32\columnwidth}
    \centering
    \includegraphics[width=\textwidth]{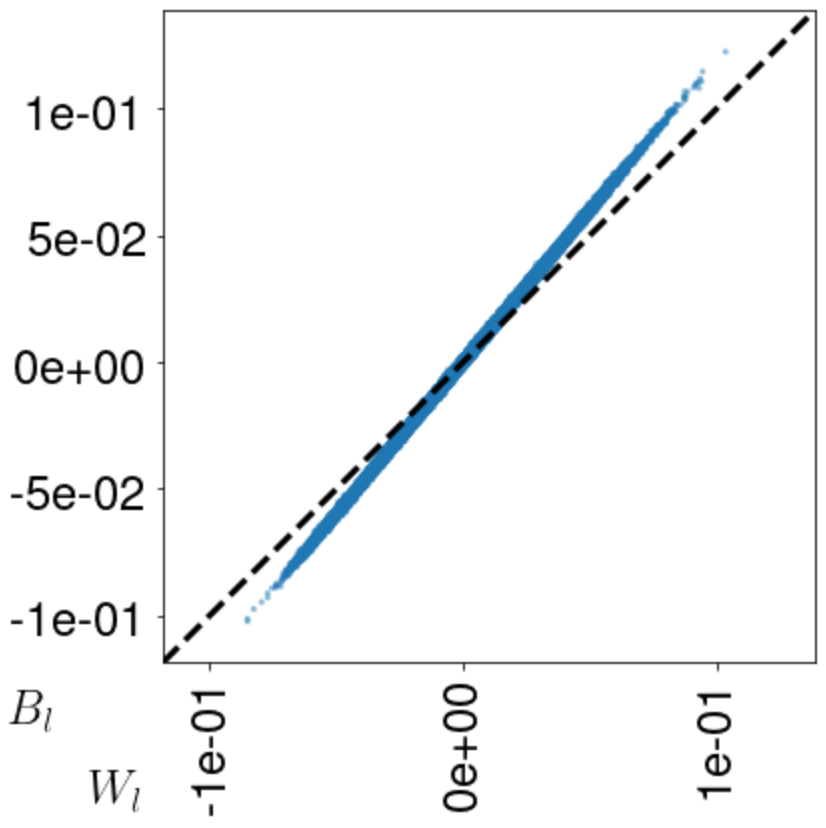}
    \caption{Conv 16}
\end{subfigure}
\begin{subfigure}{0.32\columnwidth}
    \centering
    \includegraphics[width=\textwidth]{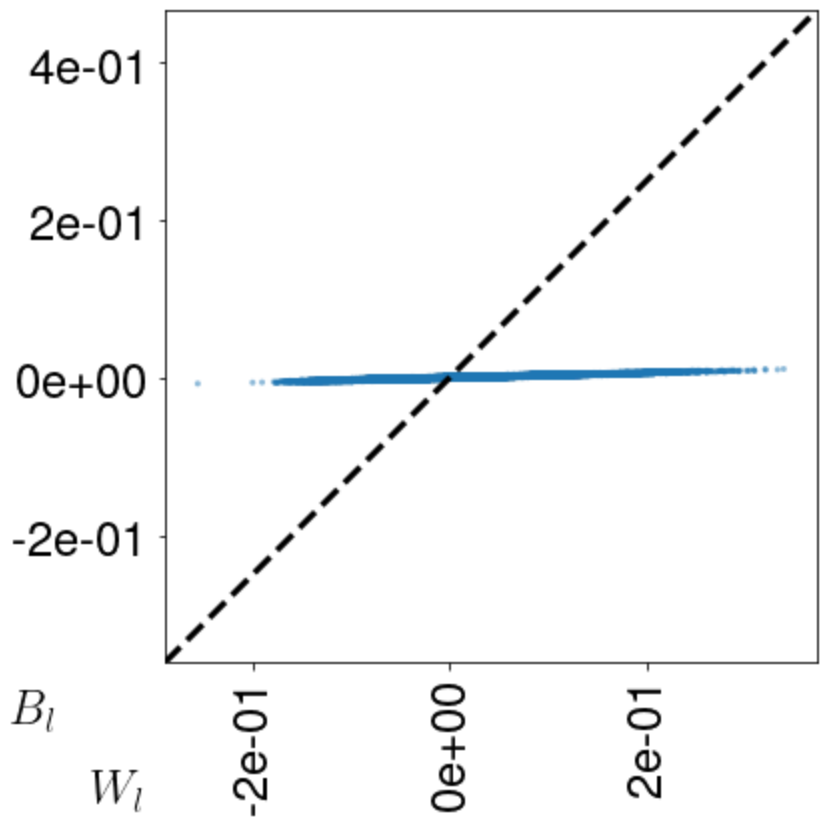}
    \caption{Dense}
\end{subfigure}
\caption{\textbf{Unstable dynamics (a-c).}  Symmetrized gradient flow on $\mathcal{R}_{\text{WM}}$ at layer $l$ with scalar weights and $x_l = 1$.  The color and arrow indicate respectively the magnitude and direction of the flow.
\textbf{Empirical instability (d-f).} Weight scatter plots for the first convolution, an intermediate convolution and the final dense layer of a ResNet-18 model trained with weight mirror for five epochs.
Each dot represents an element in layer $l$'s weight matrix and its $(x,y)$ location corresponds to its forward and backward weight values, $(W_l^{(i,j)},B_l^{(j,i)} )$. 
The dotted diagonal line shows perfect weight symmetry, as is the case in backpropagation.
Different layers demonstrate one of the the three dynamics outlined by the gradient flow analysis in \S\ref{sec:local-learning-rules}: diverging, stable, and collapsing backward weights.
\label{fig:stability}}
\vskip -0.1in
\end{figure}

In the following analysis we ignore non-linearities, include weight decay on the forward weights, set $\alpha = \beta$, and consider the gradient with respect to both the forward and backward weights.  
When the weights, $w_l$ and $b_l$, and input, $x_l$, are all scalar values, the gradient flow gives rise to the dynamical system
\begin{equation} \label{eq:dynamical_system}
\dfrac{\partial}{\partial t}\begin{bmatrix}
w_l\\
b_l
\end{bmatrix} = - A \begin{bmatrix}
w_l\\
b_l
\end{bmatrix},
\end{equation}
where $A$ is an indefinite matrix (see Appendix~\ref{sup:stability-analysis} for details.) $A$ can be diagonally decomposed by the eigenbasis $\{u,v\}$, where $u$ spans the symmetric component and $v$ spans the skew-symmetric component of any realization of the weight vector $\begin{bmatrix}
w_l &
b_l
\end{bmatrix}^\intercal$. 
Under this basis, the dynamical system decouples into a system of ODEs governed by the eigenvalues of $A$.
The eigenvalue associated with the skew-symmetric eigenvector $v$ is strictly positive, implying that this component decays exponentially to zero.
However, for the symmetric eigenvector $u$, the sign of the corresponding eigenvalue depends on the relationship between $\lambda_{\text{WM}}$ and $x_l^2$.
When $\lambda_{\text{WM}} > x_l^2$, the eigenvalue is positive and the symmetric component decays to zero (i.e. too much regularization).
When $\lambda_{\text{WM}} < x_l^2$, the eigenvalue is negative and the symmetric component exponentially grows (i.e. too little regularization).
Only when $\lambda_{\text{WM}} = x_l^2$ is the eigenvalue zero and the symmetric component stable. These various dynamics are shown in Fig.~\ref{fig:stability}.

This analysis suggests that the sensitivity of weight mirror is not due to the misalignment of the forward and backward weights, but rather due to the stability of the symmetric component throughout training.
Empirically, we find that this is true.
In Fig.~\ref{fig:stability}, we show a scatter plot of the backward and forward weights at three layers of a ResNet-18 model trained with weight mirror.
At each layer there exists a linear relationship between the weights, suggesting that the backward weights have aligned to the forward weights up to magnitude.
Despite being initialized with similar magnitudes, at the first layer the backward weights have grown orders larger, at the last layer the backward weights have decayed orders smaller, and at only one intermediate layer were the backward weights comparable to the forward weights, implying symmetry.

This analysis also clarifies the stabilizing role of Gaussian noise, which was found to be essential to weight mirror's performance gains over feedback alignment~\cite{akrout_deep_2019}. 
Specifically, when the layer input $x_l \sim N\left(0,\sigma^2\right)$ and $\sigma^2 = \lambda_{\text{WM}}$, then $x_l^2 \approx \lambda_{\text{WM}}$, implying the dynamical system in equation (\ref{eq:dynamical_system}) is stable.

\textbf{Strategies for Reducing Instability.} 
Given the above analysis, can we identify further strategies for reducing instability during learning beyond the use of Gaussian noise?

\textit{Adaptive Optimization.} 
One option is to use an adaptive learning rule strategy, such as Adam~\cite{kingma2014adam}. 
An adaptive learning rate keeps an exponentially decaying moving average of past gradients, allowing for more effective optimization of the alignment regularizer even in the presence of exploding or vanishing gradients.

\textit{Local Stabilizing Operations.} 
A second option to improve stability is to consider local layer-wise operations to the backward path such as choice of non-linear activation functions, batch centering, feature centering, and feature normalization.
The use of these operations is largely inspired by Batch Normalization \cite{ioffe2015batch} and Layer Normalization \cite{ba2016layer} which have been observed to stabilize learning dynamics.
The primary improvement that these normalizations allow for is the further conditioning of the covariance matrix at each layer's input, building on the benefits of using Gaussian noise.
In order to keep the learning rule fully local, we use these normalizations, which unlike Batch and Layer Normalization, do not add any additional learnable parameters.

\textit{The Information Alignment (IA) Learning Rule.} 
There is a third option for improving stability that involves modifying the local learning rule itself. 

Without decay, the update given by weight mirror, $\Delta B_l = \eta x_{l}x_{l+1}^\intercal$, is Hebbian.
This update, like all purely Hebbian learning rules, is unstable and can result in the norm of $B_l$ diverging.
This can be mitigated by weight decay, as is done in \citet{akrout_deep_2019}. 
However, an alternative strategy to dealing with the instability of a Hebbian update was given by \citet{oja_neuron} in his analysis of learning rules for linear neuron models.  
In the spirit of that analysis, assume that we can normalize the backward weight after each Hebbian update such that
$$B_l^{(t+1)} = \frac{B_l^{(t)} + \eta x_{l}x_{l+1}^\intercal}{||B_l^{(t)} + \eta x_{l}x_{l+1}^\intercal||},$$
and in particular $||B_l^{(t)}|| = 1$ at all time $t$. Then, for small learning rates $\eta$, the right side can be expanded as a power series in $\eta$, such that
$$B_l^{(t+1)} = B_l^{(t)} + \eta \left(x_{l}x_{l+1}^\intercal - B_l^{(t)} x_l^\intercal B_l^{(t)} x_{l+1}\right) + O(\eta^2).$$
Ignoring the $O(\eta^2)$ term gives the non-linear update
$$\Delta B_l = \eta \left( x_{l}x_{l+1}^\intercal - B_l x_{l}^\intercal B_l x_{l+1}\right).$$
If we assume $x_l^\intercal B_l = x_{l+1}$ and $B_l$ is a column vector rather than a matrix, then by Table~\ref{tab:prim}, this is approximately the update given by the null primitive introduced in \S\ref{sec:framework-primitives}.

Thus motivated, we define \textbf{Information Alignment (IA)} as the local learning rule defined by adding a (weighted) null primitive to the other two local primitives already present in the weight mirror rule. 
That is, the layer-wise regularization function
$$\mathcal{R}_{\text{IA}} = \sum_{l \in \text{layers}}\alpha\mathcal{P}^{\text{amp}}_l +  \beta\mathcal{P}^{\text{decay}}_l + \gamma\mathcal{P}^{\text{null}}_l.$$
In the specific setting when $x_{l+1} = W_lx_l$ and $\alpha = \gamma$, then the gradient of $\mathcal{R}_{\text{IA}}$ is proportional to the gradient with respect to $B_l$ of
$\frac{1}{2}||x_l - B_lW_lx_l||^2 + \frac{\beta}{2}\left(||W_l||^2 + ||B_l||^2\right)$, a quadratically regularized linear autoencoder\footnote{In this setting, the resulting learning rule is a member of the target propagation framework introduced in \S\ref{sec:related}.}.
As shown in \citet{kunin_loss_2019}, all critical points of a quadratically regularized linear autoencoder attain symmetry of the encoder and decoder.
% \footnote{Further, when $B_l$ is the pseudoinverse of $W_l$ at all layers $l$, then the updates generated by the backward pass are approximately the updates given by Gauss-Newton optimization \citep{lillicrap_random_2016}.}

\begin{table}
    \setlength\tabcolsep{3pt}
    \centering
    \begin{tabular}{ccc}
    \toprule
    Learning Rule & Top-1 Val Accuracy & Top-5 Val Accuracy\\
    \midrule
     $\mathcal{R}_{\text{WM}}$ & 63.5\% & 85.16\% \\
    \midrule
     $\mathcal{R}_{\text{WM}}^{\text{TPE}}$ & 64.07\% & 85.47\% \\
     \midrule
     $\mathcal{R}_{\text{WM}+\text{AD}}^{\text{TPE}}$ & 64.40\% & 85.53\% \\
     \midrule
     $\mathcal{R}_{\text{WM}+\text{AD}+\text{OPS}}^{\text{TPE}}$ & 63.41\% & 84.83\% \\
     \midrule
     $\mathcal{R}_{\text{IA}}^{\text{TPE}}$ & \textbf{67.93\%} & \textbf{88.09\%} \\
    \midrule
    \text{Backprop.} & 70.06\% & 89.14\%\\
    \bottomrule
    \end{tabular}
    \caption{\textbf{Performance of local learning rules with ResNet-18 on ImageNet.} $\mathcal{R}_{\text{WM}}$ is weight mirror as described in \citet{akrout_deep_2019}, $\mathcal{R}_{\text{WM}}^{\text{TPE}}$ is weight mirror with learning metaparameters chosen through an optimization procedure.  $\mathcal{R}_{\text{WM}+\text{AD}}^{\text{TPE}}$ is weight mirror with an adaptive optimizer. $\mathcal{R}_{\text{WM}+\text{AD}+\text{OPS}}^{\text{TPE}}$ involves the addition of stabilizing operations to the network architecture. The best local learning rule, $\mathcal{R}_{\text{IA}}^{\text{TPE}}$, additionally involves the null primitive.  
    For details on metaparameters for each local rule, see Appendix~\ref{sup:hp-ss-details}.}
    \label{tab:hp-local}
    \vskip -0.2in
\end{table}

\textbf{Empirical Results.} 
To evaluate the three strategies for stabilizing local weight updates, we performed a neural architecture search implementing all three strategies, again using TPE.  
This search optimized for Top-1 ImageNet validation performance with the ResNet-18 architecture, comprising a total of 628 distinct settings.   
We found that validation performance increased significantly, with the optimal learning rule $\mathcal{R}_{\text{IA}}^{\text{TPE}}$, attaining 67.93\% top-1 accuracy (Table \ref{tab:hp-local}). 
More importantly, we also found that the parameter robustness of $\mathcal{R}_{\text{IA}}^{\text{TPE}}$ is dramatically improved as compared to weight mirror (Fig. \ref{fig:hp-deeper}, orange line), nearly equaling the parameter robustness of backpropagation across a variety of deeper architectures.  
Critically, this improvement was achieved not by directly optimizing for robustness across architectures, but simply by finding a parameter setting that achieved high task performance on one architecture.

To assess the importance of each strategy type in achieving this result, we also performed several ablation studies, involving neural architecture searches using only various subsets of the stabilization strategies (see Appendix~\ref{sup:hp-wm-ad-ops-tpe-details} for details).
Using just the adaptive optimizer while otherwise optimizing the weight mirror metaparameters yielded the learning rule $\mathcal{R}_{\text{WM}+\text{AD}}^{\text{TPE}}$, while adding stabilizing layer-wise operations yielded the learning rule $\mathcal{R}_{\text{WM}+\text{AD}+\text{OPS}}^{\text{TPE}}$ (Table \ref{tab:hp-local}).  
We found that while the top-1 performance of these ablated learning rules was not better for the ResNet-18 architecture than the weight-mirror baseline, each of the strategies did individually contribute significantly to improved parameter robustness (Fig. \ref{fig:hp-deeper}, red and green lines). 

Taken together, these results indicate that the regularization framework allows the formulation of local learning rules with substantially improved performance and, especially, metaparameter robustness characteristics. 
Moreover, these improvements are well-motivated by mathematical analysis that indicates how to target better circuit structure via improved learning stability.

\section{Non-Local Learning Rules}
\label{sec:non-local-learning-rules}
While our best local learning rule is substantially improved as compared to previous alternatives, it still does not quite match backpropagation, either in terms of performance or metaparameter stability over widely different architectures (see the gap between blue and orange lines in Fig. \ref{fig:hp-deeper}). 
We next introduce two novel non-local learning rules that entirely eliminate this gap.

\textbf{Symmetric Alignment (SA)} is defined by the layer-wise regularization function
$$\mathcal{R}_{\text{SA}} = \sum_{l \in \text{layers}} \alpha\mathcal{P}^{\text{self}}_l + \beta\mathcal{P}^{\text{decay}}_l.$$
When $\alpha=\beta$, then the gradient of $\mathcal{R}_{\text{SA}}$ is proportional to the gradient with respect to $B_l$ of $\frac{1}{2}||W_l - B_l^\intercal ||^2$, which encourages symmetry of the weights.

\textbf{Activation Alignment (AA)} is defined by the layer-wise regularization function
$$\mathcal{R}_{\text{AA}} = \sum_{l \in \text{layers}} \alpha\mathcal{P}^{\text{amp}}_l + \beta\mathcal{P}^{\text{sparse}}_l.$$
When $x_{l+1} = W_lx_l$ and $\alpha = \beta$, then the gradient of $\mathcal{R}_{\text{AA}}$ is proportional to the gradient with respect to $B_l$ of $\frac{1}{2}||W_lx_l - B_l^\intercal x_l||^2$, which encourages alignment of the activations.

Both SA and AA give rise to dynamics that encourage the backward weights to become transposes of their forward counterparts.
When $B_l$ is the transpose of $W_l$ for all layers $l$ then the updates generated by the backward pass are the exact gradients of $\mathcal{J}$.
It follows intuitively that throughout training the pseudogradients given by these learning rules might converge to better approximations of the exact gradients of $\mathcal{J}$, leading to improved learning.
Further, in the context of the analysis in equation (\ref{eq:dynamical_system}), the matrix $A$ associated with SA and AA is positive semi-definite, and unlike the case of weight mirror, the eigenvalue associated with the symmetric eigenvector $u$ is zero, implying stability of the symmetric component.

While weight mirror and Information Alignment introduce dynamics that implicitly encourage symmetry of the forward and backward weights, the dynamics introduced by SA and AA encourage this property explicitly.

Despite not having the desirable locality property, we show that SA and AA perform well empirically in the weight-decoupled regularization framework --- meaning that they \emph{do} relieve the need for exact weight symmetry.  
As we will discuss, this may make it possible to find plausible biophysical mechanisms by which they might be implemented.

\textbf{Parameter Robustness of Non-Local Learning Rules.}
To assess the robustness of SA and AA, we trained ResNet-18 models with standard 224-sized ImageNet images (training details can be found in Appendix~\ref{sup:non-local-hp}).
Without any metaparameter tuning, SA and AA were able to match backpropagation in performance. 
Importantly, for SA we did not need to employ any specialized or adaptive learning schedule involving alternating modes of learning, as was required for all the local rules.
However, for AA we did find it necessary to use an adaptive optimizer when minimizing $\mathcal{R}_{\text{AA}}$, potentially due to the fact that it appears to align less exactly than SA (see Fig.~\ref{fig:full_weight_scatter}).
We trained deeper ResNet-50, 101, and 152 models \cite{he_deep_2016} with larger 299-sized ImageNet images. 
As can be seen in Table~\ref{tab:empimnet}, both SA and AA maintain consistent performance with backpropagation despite changes in image size and increasing depth of network architecture, demonstrating their robustness as a credit assignment strategies.

\begin{table}
\centering
\begin{tabular}{cccc}
\toprule
Model & Backprop. & Symmetric & Activation \\
\midrule
ResNet-18 & 70.06\% & 69.84\% & 69.98\% \\
\midrule
ResNet-50 & 76.05\% & 76.29\% & 75.75\% \\
\midrule
ResNet-50v2 & 77.21\% & 77.18\% & 76.67\% \\
\midrule
ResNet-101v2 & 78.64\% & 78.74\% & 78.35\% \\
\midrule
ResNet-152v2 & 79.31\% & 79.15\% & 78.98\% \\
\bottomrule
\end{tabular}
\caption{\textbf{Symmetric and Activation Alignment consistently match backpropagation.}
Top-1 validation accuracies on ImageNet for each model class and non-local learning rule, compared to backpropagation.
\label{tab:empimnet}}
\vskip -0.3in
\end{table}

\begin{figure}[h]
\vskip 3mm
    \centering
    \begin{subfigure}{\columnwidth}
        \centering
        \includegraphics[width=0.85\textwidth]{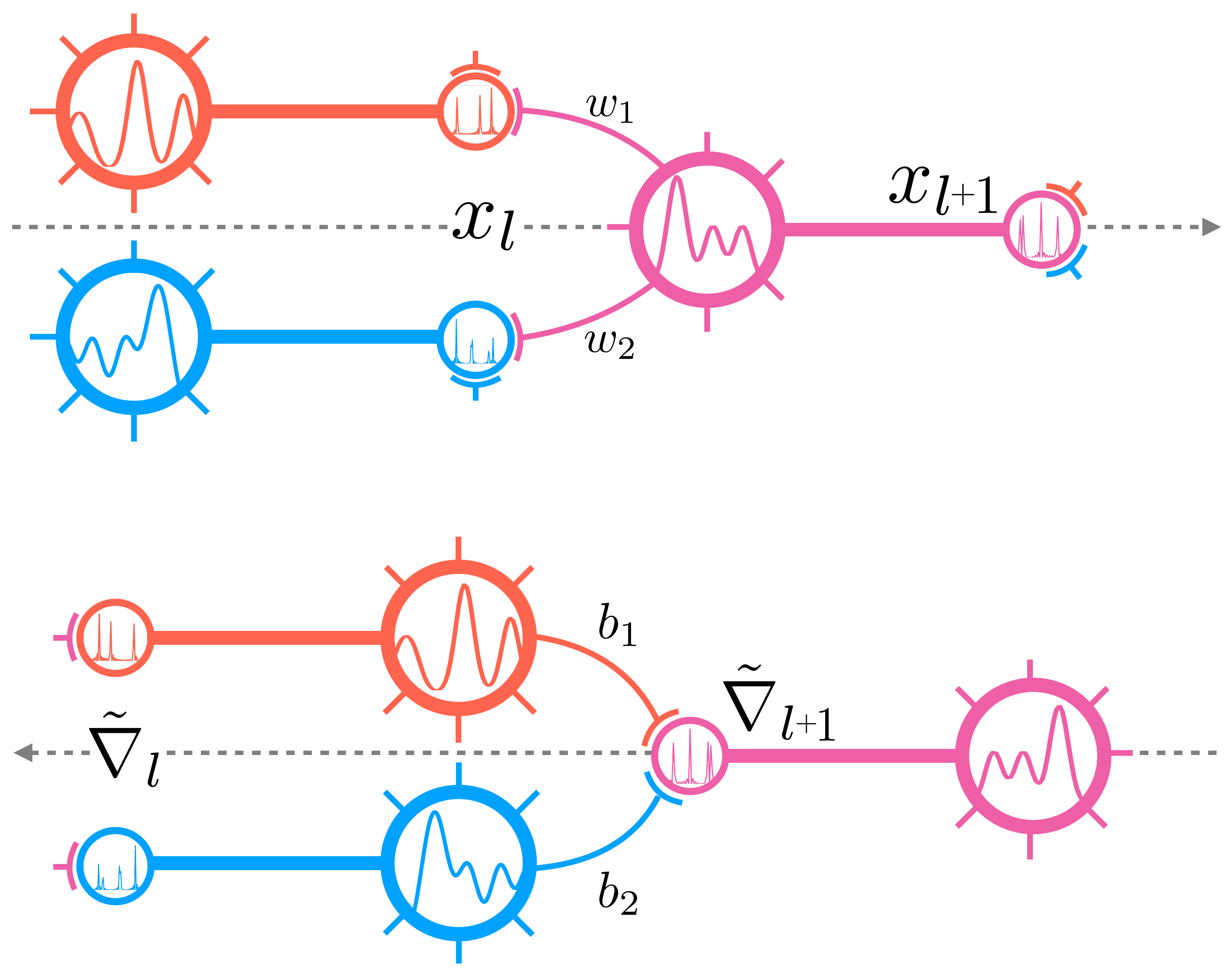}
        \caption{Neural Circuit Diagram}
    \end{subfigure}
\vskip 2mm
    \begin{subfigure}{\columnwidth}
        \centering
        \includegraphics[width=0.49\textwidth]{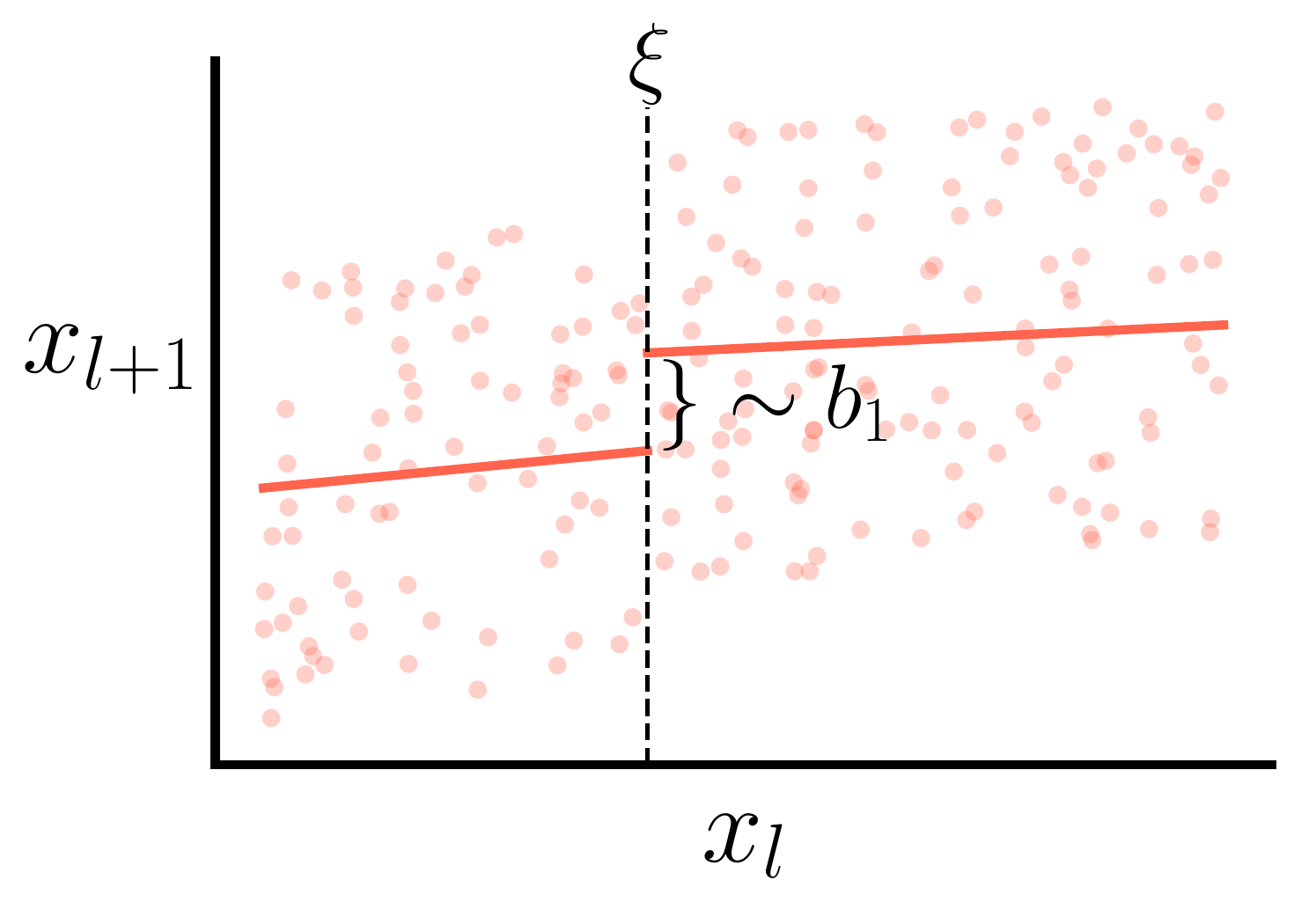}
        \includegraphics[width=0.49\textwidth]{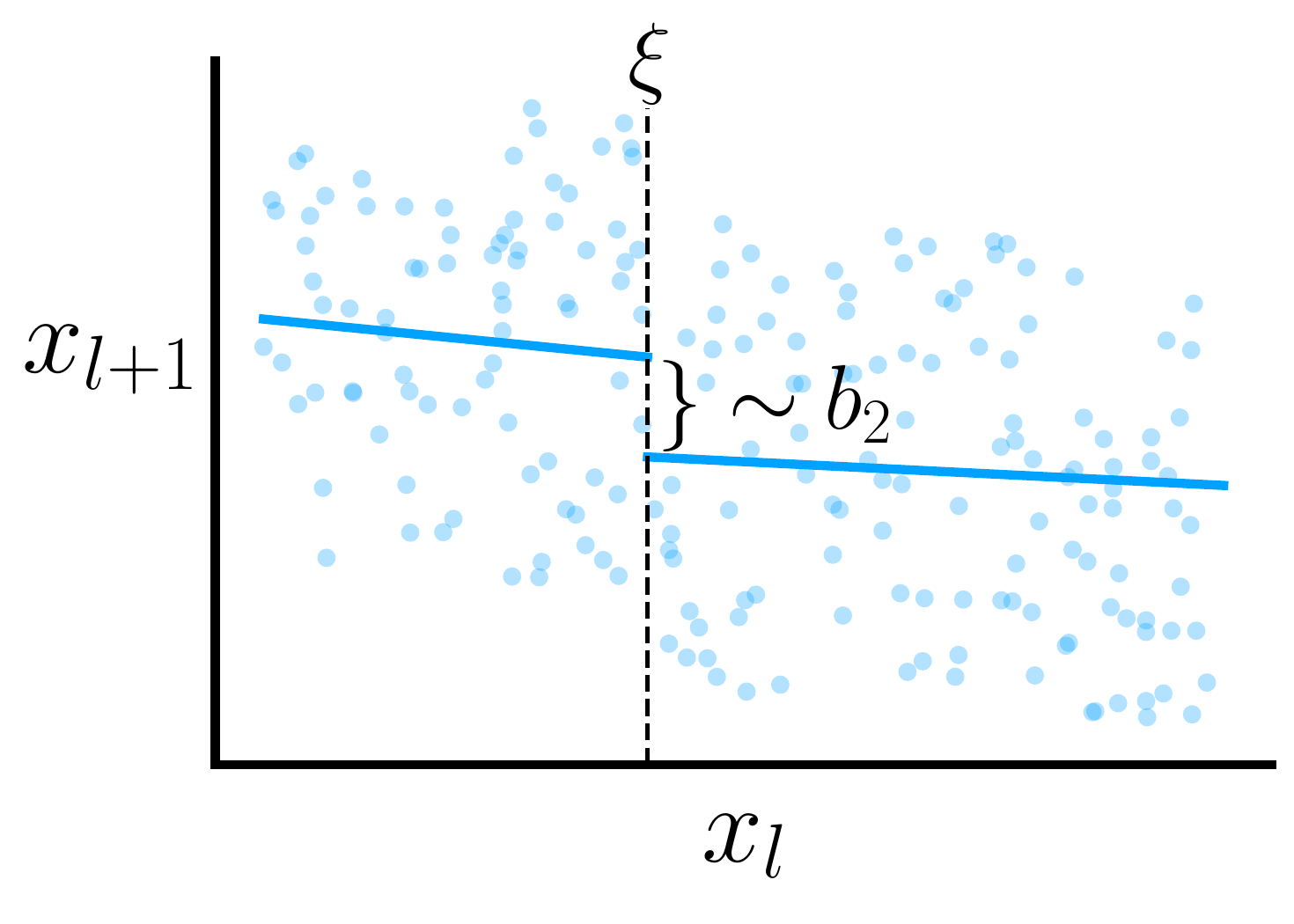}
        \caption{Regression Discontinuity Design}
    \end{subfigure}
    \caption{\textbf{Weight estimation.} (a) The notational diagram, as shown in Fig.~\ref{fig:conceptual-framework}, is mapped into a neural circuit.
    The top and bottom circuits represent the forward and backward paths respectively.
    Large circles represent the soma of a neuron, thick edges the axon, small circles the axon terminal, and thin edges the dendrites.
    Dendrites corresponding to the lateral pathways between the circuits are omitted.
    (b) A mechanism such as regression discontinuity design, as explained by \citet{lansdell2019spiking} and \citet{guerguiev_spike-based_2019}, could be used independently at each neuron to do weight estimation by quantifying the causal effect of $x_l$ on $x_{l+1}$.}
    \label{fig:rdd_neurons}
    \vskip -0.2in
\end{figure}

\textbf{Weight Estimation, Neural Plausibility, and Noise Robustness.}
Though SA is non-local, it does avoid the need for instantaneous weight transport --- as is shown simply by the fact that it optimizes effectively in the framework of decoupled forward-backward weight updates, where alignment can only arise over time due to the structure of the regularization circuit rather than instantaneously by \emph{fiat} at each timepoint.
Because of this key difference, it may be possible to find plausible biological implementations for SA, operating on a principle of iterative ``weight estimation'' in place of the implausible idea of instantaneous weight transport.

By ``weight estimation'' we mean any process that can measure changes in post-synaptic activity relative to varying synaptic input, thereby providing a temporal estimate of the synaptic strengths.
Prior work has shown how noisy perturbations in the presence of spiking discontinuities \cite{lansdell2019spiking} could provide neural mechanisms for weight estimation, as depicted in Fig.~\ref{fig:rdd_neurons}.
In particular, \citet{guerguiev_spike-based_2019} present a spiking-level mechanism for estimating forward weights from noisy dendritic measurements of the implied effect of those weights on activation changes. 
This idea, borrowed from the econometrics literature, is known as regression discontinuity design \cite{imbens2008regression}. 
This is essentially a form of iterative weight estimation, and is used in \citet{guerguiev_spike-based_2019} for minimizing a term that is mathematically equivalent to $\mathcal{P}^{\text{self}}$.
\citet{guerguiev_spike-based_2019} demonstrate that this weight estimation mechanism works empirically for small-scale networks.

\begin{figure}[t!]
\centering
\begin{subfigure}{0.9\columnwidth}
    \centering
    \includegraphics[width=\textwidth]{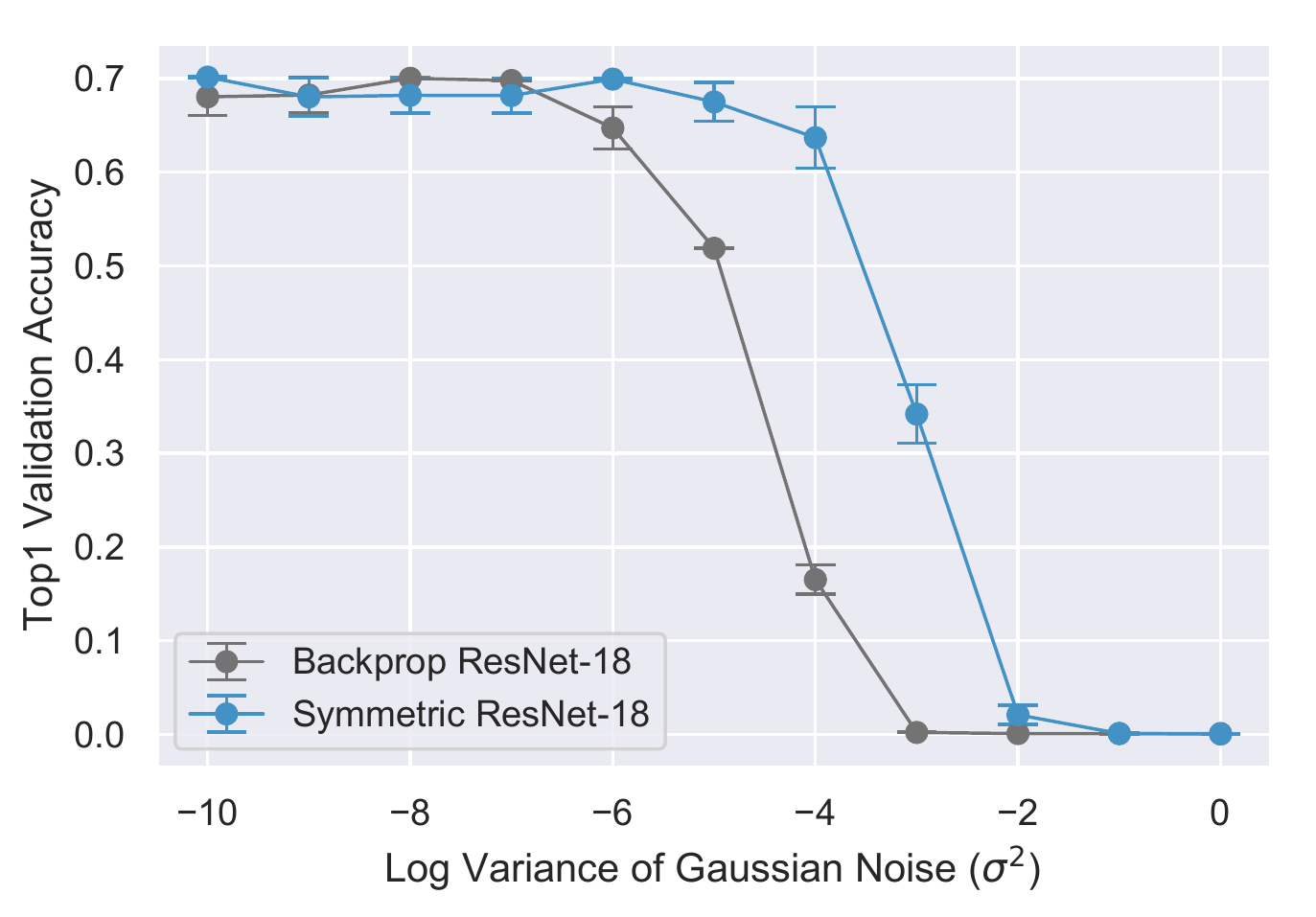}
    \caption{ResNet-18}
    \label{fig:noise-plot-a}
\end{subfigure}
\begin{subfigure}{0.9\columnwidth}
    \centering
    \includegraphics[width=\textwidth]{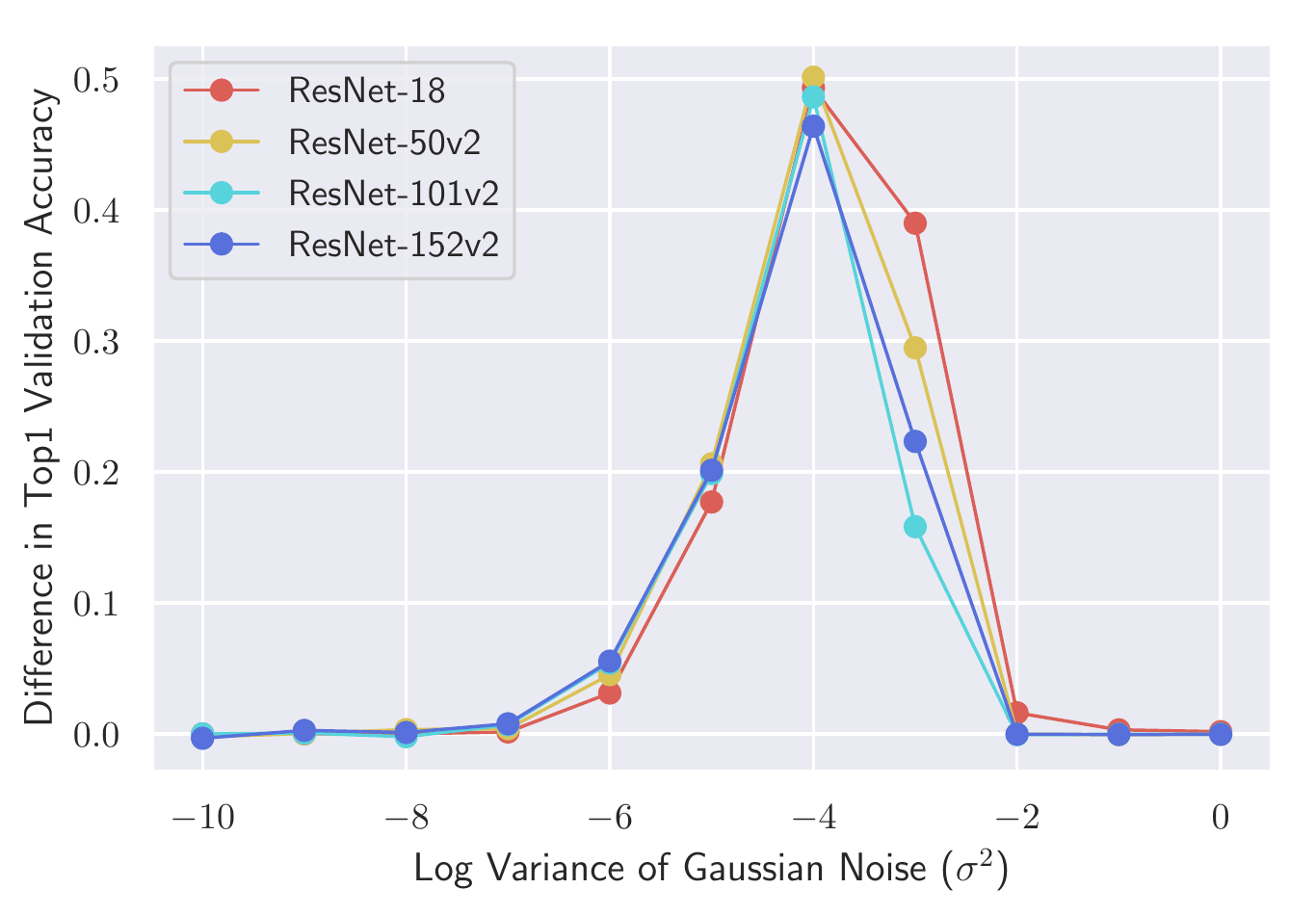}
    \caption{Deeper Models}
    \label{fig:noise-plot-b}
\end{subfigure}
\caption{\textbf{Symmetric Alignment is more robust to noisy updates than backpropagation.} (a) Symmetric Alignment is more robust than backpropagation to increasing levels of Gaussian noise added to its updates for ResNet-18. 
(b) Symmetric Alignment maintains this robustness for deeper models.
See Appendix~\ref{sup:noisy-updates} for more details and similar experiments with Activation Alignment.}
\label{fig:noise-plot}
\vskip -0.15in
\end{figure}

Our performance and robustness results above for SA can be interpreted as providing evidence that a rate-coded version of weight estimation scales effectively to training deep networks on large-scale tasks.
However, there remains a gap between what we have shown at the rate-code level and the spike level, at which the weight estimation mechanism operates.
Truly showing that weight estimation could work at scale would involve being able to train deep spiking neural networks, an unsolved problem that is beyond the scope of this work. 
One key difference between any weight estimation process at the rate-code and spike levels is that the latter will be inherently noisier due to statistical fluctuations in whatever local measurement process is invoked --- e.g. in the \citet{guerguiev_spike-based_2019} mechanism, the noise in computing the regression discontinuity.

As a proxy to better determine if our conclusions about the scalable robustness of rate-coded SA are likely to apply to spiking-level equivalents, we model this uncertainty by adding Gaussian noise to the backward updates during learning.
To the extent that rate-coded SA is robust to such noise, the more likely it is that a spiking-based implementation will have the performance and parameter robustness characteristics of the rate-coded version. 
Specifically, we modify the update rule as follows:
\vskip -0.2in
\begin{equation*}
\Delta \theta_b \propto \nabla \mathcal{R} +  \mathcal{N}(0,\sigma^2),\qquad \Delta \theta_f \propto \widetilde{\nabla}\mathcal{J}.
\end{equation*}
As shown in Fig.~\ref{fig:noise-plot}, the performance of SA is very robust to noisy updates for training ResNet-18.
In fact, for comparison we also train backpropagation with Gaussian noise added to its gradients,
$\Delta \theta \propto \nabla \mathcal{J} + \mathcal{N}(0,\sigma^2)$, and find that SA is substantially \emph{more} robust than backpropagation.
For deeper models, SA maintains this robustness, implying that pseudogradients generated by backward weights with noisy updates leads to more robust learning than using equivalently noisy gradients directly.

\section{Discussion}
\label{sec:discussion}

\begin{figure}[tb]
\centering
\includegraphics[width=0.8\columnwidth]{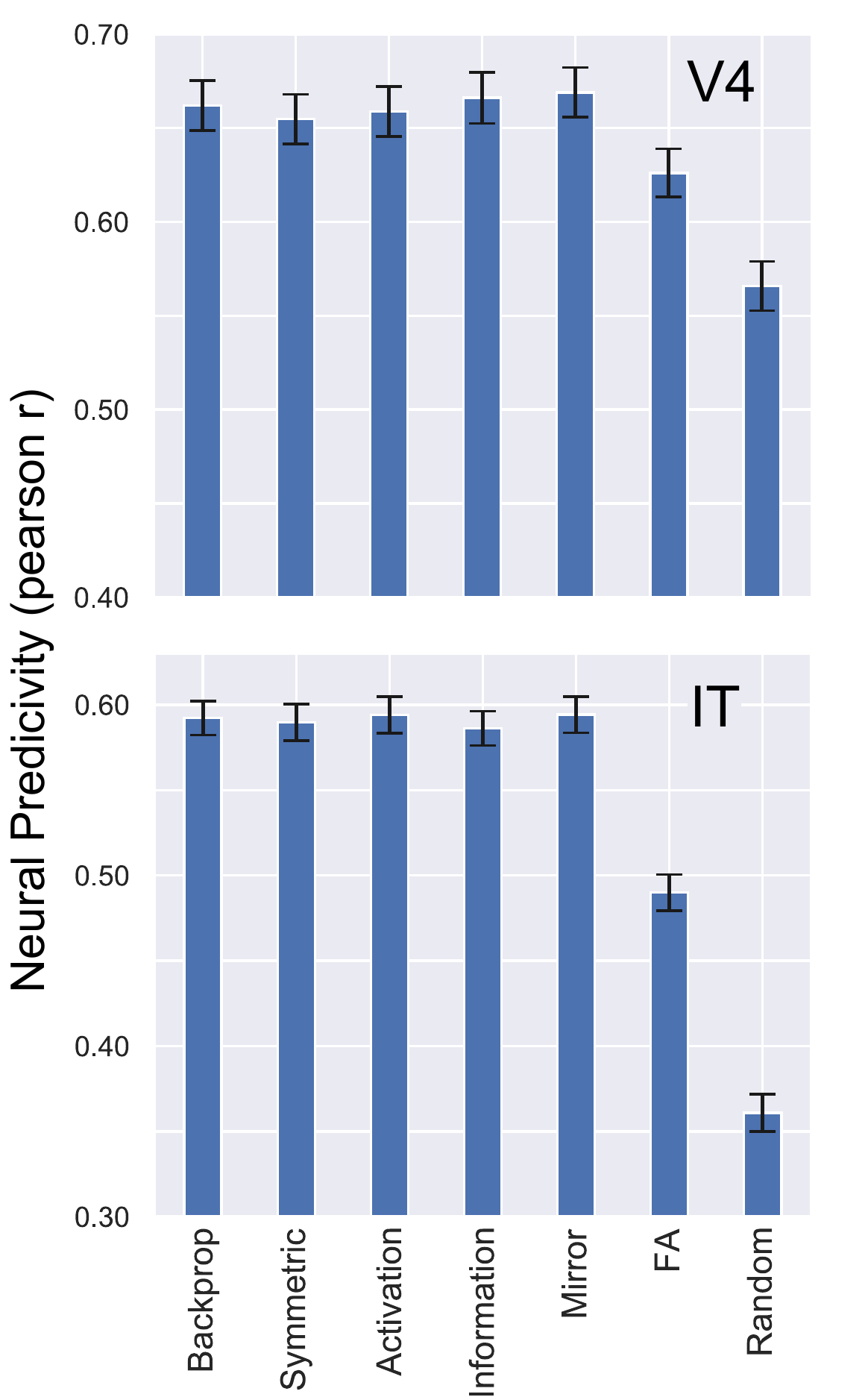}
\caption{
\textbf{Neural fits to temporally-averaged V4 and IT responses.}
Neural fits to V4 (top) and IT (bottom) time-averaged responses \cite{majaj2015simple}, using a 25 component PLS mapping on a ResNet-18.
The median (across neurons) Pearson correlation over images, with standard error of mean (across neurons) denoting the error bars.
``Random'' refers to a ResNet-18 architecture at initialization. 
For details, see Appendix~\ref{sup:neural-fit}.
\label{fig:neural-fits}}
\vskip -0.1in
\end{figure}

In this work, we present a unifying framework that allows for the systematic identification of robust and scalable alternatives to backpropagation.
We obtain, through large-scale searches, a local learning rule that transfers more robustly across architectures than previous local alternatives.
Nonetheless, a performance and robustness gap persists with backpropagation.
We formulate non-local learning rules that achieve competitive performance with backpropagation, requiring almost no metaparameter tuning and are robust to noisy updates.
Taken together, our findings suggest that there are two routes towards the discovery of robust, scalable, and neurally plausible credit assignment without weight symmetry.

The first route involves further improving local rules. 
We found that the local operations and regularization primitives that allow for improved approximation of non-local rules perform better and are much more stable. 
If the analyses that inspired this improvement could be refined, perhaps further stability could be obtained. 
To aid in this exploration going forward, we have written an open-source TensorFlow library\footnote{\url{https://github.com/neuroailab/neural-alignment}}, allowing others to train arbitrary network architectures and learning rules at scale, distributed across multiple GPU or TPU workers.
The second route involves the further refinement and characterization of scalable biological implementations of weight estimation mechanisms for Symmetric or Activation Alignment, as \citet{guerguiev_spike-based_2019} initiate.

Given these two routes towards neurally-plausible credit assignment without weight symmetry, how would we use neuroscience data to adjudicate between them?
It would be convenient if functional response data in a ``fully trained'' adult animal showed a signature of the underlying learning rule, without having to directly measure synaptic weights during learning.
Such data have been very effective in identifying good models of the primate ventral visual pathway~\cite{majaj2015simple, yamins2014performance}. 
As an initial investigation of this idea, we compared the activation patterns generated by networks trained with each local and non-local learning rule explored here, to neural response data from several macaque visual cortical areas, using a regression procedure similar to that in \citet{yamins2014performance}.
As shown in Fig.~\ref{fig:neural-fits}, we found that, with the exception of the very poorly performing feedback alignment rule, all the reasonably effective learning rules achieve similar V4 and IT neural response predictivity, and in fact match that of the network learned via backpropagation. 
Such a result suggests the interesting possibility that the functional response signatures in an already well-learned neural representation may be relatively independent of which learning rule created them. 
Perhaps unsurprisingly, the question of identifying the operation of learning rules in an \emph{in vivo} neural circuit will likely require the deployment of more sophisticated neuroscience techniques.

\section*{Acknowledgements}
We thank the Stanford Data Science Scholars program (DK), the Burroughs Wellcome (DY), Simons (SG, DY) and James S. McDonnell (DY) foundations, NSF career awards (SG, DY), and the NSF Robust Intelligence program, for support.
We thank the Google TensorFlow Research Cloud (TFRC) team for providing TPU resources for this project.

% References
\bibliography{main}
\bibliographystyle{icml2020}

%%%%%%%%%%%%%%%%%%%%%%%%%%%%%%%%%%%%%%%%%%%%%%%%%%%%%%%%%%%%%%%%%%%%%%%%%%%%%%%
%%%%%%%%%%%%%%%%%%%%%%%%%%%%%%%%%%%%%%%%%%%%%%%%%%%%%%%%%%%%%%%%%%%%%%%%%%%%%%%
% DO NOT PLACE CONTENT AFTER THE REFERENCES! (This is only for ICML)
%%%%%%%%%%%%%%%%%%%%%%%%%%%%%%%%%%%%%%%%%%%%%%%%%%%%%%%%%%%%%%%%%%%%%%%%%%%%%%%
%%%%%%%%%%%%%%%%%%%%%%%%%%%%%%%%%%%%%%%%%%%%%%%%%%%%%%%%%%%%%%%%%%%%%%%%%%%%%%%

\newpage
\appendix
\setcounter{figure}{0}
\renewcommand{\thefigure}{S\arabic{figure}}

\section{Code base}
In this section we describe our implementation and highlight the technical details that allow its generality for use in any architecture. 
We used TensorFlow version 1.13.1 to conduct all experiments, and adhered to its interface. All code can be found at \url{https://github.com/neuroailab/neural-alignment}.

\subsection{Layers}
The essential idea of our code base is that by implementing custom layers that match the TensorFlow API, but use custom operations for matrix multiplication (\texttt{matmul}) and two-dimensional convolutions (\texttt{conv2d}), then we can efficiently implement arbitrary feedforward networks using any credit assignment strategies with untied forward and backward weights.
Our custom \texttt{matmul} and \texttt{conv2d} operations take in a forward and backward kernel. 
They use the forward kernel for the forward pass, but use the backward kernel when propagating the gradient.
To implement this, we leverage the \texttt{@tf.custom\_gradient} decorator, which allows us to explicitly define the forward and backward passes for that op. 
Our \texttt{Layer} objects implement custom dense and convolutional layers which use the custom operations described above. 
Both layers take in the same arguments as the native TensorFlow layers and an additional argument for the learning rule.

\subsection{Alignments}
A learning rule is defined by the form of the layer-wise regularization $\mathcal{R}$ added to the model at each layer.  
The custom layers take an instance of an alignment class which when called will define its alignment specific regularization and add it to the computational graph. 

The learning rule are specializations of a parent \texttt{Alignment} object which implements a \texttt{\_\_call\_\_} method that creates the regularization function. 
The regularization function uses tensors that prevent the gradients from flowing to previous layers via \texttt{tf.stop\_gradient}, keeping the alignment loss localized to a single layer.
Implementation of the \texttt{\_\_call\_\_} method is delegated to subclasses, such that they can define their alignment specific regularization as a weighted sum of primitives, each of which is defined as a function. 

% Thus, our custom layers in addition to accepting the TensorFlow arguments, they accept an \texttt{alignment} argument, which should be an instance a specific subclass described above. 
% The alignment's \texttt{\_\_call\_\_} method is then called within the layer, to add the appropriate regularization. 

\subsection{Optimizers}
The total network loss is defined as the sum of the global cost function $\mathcal{J}$ and the local alignment regularization $\mathcal{R}$.
The optimizer class provides a framework for specifying how to optimize each part of the total network loss as a function of the global step.

In the \texttt{Optimizers} folder you will find two important files:
\begin{itemize}
    \item \texttt{rate\_scheduler.py} defines a scheduler which is a function of the global step, that allows you to adapt the components of the alignment weighting based on where it is in training. If you do not pass in a scheduling function, it will by default return a constant rate.
    \item \texttt{optimizers.py} provides a class which takes in a list of optimizers, as well as a list of losses to optimize. Each loss element is optimized with the corresponding optimizer at each step in training, allowing you to have potentially different learning rate schedules for different components of the loss. Minibatching is also supported.
\end{itemize}

\section{Experimental Details}

\begin{table*}[t]
\resizebox{\textwidth}{!}{
    \begin{tabular}{@{}rrrcrrcrrcrr@{}}
    \toprule
                        & $\mathcal{R}_{\text{WM}}^{\text{TPE}}$ & $\mathcal{R}_{\text{WM}+\text{AD}}^{\text{TPE}}$ & $\mathcal{R}_{\text{WM}+\text{AD}+\text{OPS}}^{\text{TPE}}$ & $\mathcal{R}_{\text{IA}}^{\text{TPE}}$ \\ \midrule
    Alternating Minimization & True & True & True & True \\
    Delay Epochs ($\text{de}$) & 2 & 0 & 0 & 1 \\
    Train Batch Size ($|\mathcal{B}|$) & 2048 & 256 & 256 & 256 \\
    SGDM Learning Rate & 1.0 & 0.125 & 0.125 & 0.125 \\
    Alignment Optimizer & Vanilla GD & Adam & Adam & Adam \\
    Alignment Learning Rate ($\eta$) & 1.0 & 0.0053 & 0.0025 & 0.0098 \\
    $\sigma$ & 0.6905 & 0.9500 & 0.6402 & 0.8176 \\
    $\alpha/\beta$ & 15.6607 & 13.9040 & 0.1344 & 129.1123 \\
    $\beta$ & 0.0283 & $2.8109 \times 10^{-8}$ & 232.7856 & 7.9990 \\
    $\gamma$ & N/A & N/A & N/A & $3.1610 \times 10^{-6}$ \\
    Forward Path Output (FO) Bias & True & True & True & True \\
    FO ReLU & True & True & True & True \\
    FO BWMC & True & True & True & True \\
    FO FWMC & False & False & True & True \\
    FO FWL2N & False & False & False & False \\
    Backward Path Output (BO) Bias & False & False & False & True \\
    BO ReLU & False & False & True & False \\
    BO FWMC & False & False & False & True \\
    BO FWL2N & False & False & True & True \\
    Backward Path Input (BI) BWMC & True & True & True & False \\
    BI FWMC & False & False & False & False \\
    BI FWL2N & False & False & False & False \\
    \bottomrule
    \end{tabular}
}
\caption{Metaparameter settings (rows) for each of the learning rules obtained by large-scale searches (columns). Continuous values were rounded up to 4 decimal places.}
\label{tab:metaparameters}
\end{table*}

In what follows we describe the metaparameters we used to run each of the experiments reported above, tabulated in Table~\ref{tab:metaparameters}.
Any defaults from TensorFlow correspond to those in version 1.13.1.

\subsection{TPE search spaces}
\label{sup:hp-ss-details}
We detail the search spaces for each of the searches performed in \S\ref{sec:local-learning-rules}.
For each search, we trained approximately 60 distinct settings at a time using the HyperOpt package \cite{bergstra_tpe_2011} using the ResNet-18 architecture and L2 weight decay of $\lambda = 10^{-4}$ \cite{he_deep_2016} for 45 epochs, corresponding to the point in training midway between the first and second learning rate drops. 
Each model was trained on its own Tensor Processing Unit (TPUv2-8 and TPUv3-8).

We employed a form of Bayesian optimization, a Tree-structured Parzen Estimator (TPE), to search the space of continuous and categorical metaparameters \cite{bergstra_tpe_2011}. This algorithm constructs a generative model of $P[score\mid configuration]$ by updating a prior from a maintained history $H$ of metaparameter configuration-loss pairs. The fitness function that is optimized over models is the expected improvement, where a given configuration $c$ is meant to optimize $EI(c) = \int_{x < t}P[x\mid c, H]$. This choice of Bayesian optimization algorithm models $P[c\mid x]$ via a Gaussian mixture, and restricts us to tree-structured configuration spaces.

\subsubsection{$\mathcal{R}_{\text{WM}}^{\text{TPE}}$ search space}
\label{sup:hp-wm-tpe-details}
Below is a description of the metaparameters and their ranges for the search that gave rise to $\mathcal{R}_{\text{WM}}^{\text{TPE}}$ in Table~\ref{tab:hp-local}.

\begin{itemize}
\item Gaussian input noise standard deviation $\sigma \in [10^{-10}, 1]$ used in the backward pass, sampled uniformly.
%\textbf{$\mathcal{R}_{\text{WM}}^{\text{TPE}}$ ended up setting $\mathbf{\sigma = 0.690465059717068}$.}

\item Ratio between the weighting of $\mathcal{P}^{\text{amp}}$ and $\mathcal{P}^{\text{decay}}$ given by $\alpha/\beta \in [0.1, 200]$, sampled uniformly. 
%\textbf{$\mathcal{R}_{\text{WM}}^{\text{TPE}}$ ended up setting $\mathbf{\alpha/\beta = 15.6606919586}$.}

\item The weighting of $\mathcal{P}^{\text{decay}}$ given by $\beta \in [10^{-11}, 10^{7}]$, sampled log-uniformly.
% \textbf{$\mathcal{R}_{\text{WM}}^{\text{TPE}}$ ended up setting $\mathbf{\beta = 0.028327320537228435}$.}
\end{itemize}
We fix all other metaparameters as prescribed by \citet{akrout_deep_2019}, namely batch centering the backward path inputs and forward path outputs in the backward pass, as well as applying a ReLU activation function and bias to the forward path but not to the backward path in the backward pass.
To keep the learning rule fully local, we do not allow for any transport during the mirroring phase of the batch normalization mean and standard deviation as \citet{akrout_deep_2019} allow.

\subsubsection{$\mathcal{R}_{\text{WM}+\text{AD}}^{\text{TPE}}$ search space}
\label{sup:hp-wm-ad-tpe-details}
Below is a description of the metaparameters and their ranges for the search that gave rise to $\mathcal{R}_{\text{WM}+\text{AD}}^{\text{TPE}}$ in Table~\ref{tab:hp-local}.
\begin{itemize}
\item Train batch size $|\mathcal{B}| \in \{256, 1024, 2048, 4096\}$.  
%\textbf{$\mathcal{R}_{\text{WM}+\text{AD}}^{\text{TPE}}$ ended up setting $\mathbf{|\mathcal{B}| = 256}$.} 
This choice also determines the forward path Nesterov momentum learning rate on the \emph{pseudogradient} of the categorization objective $\mathcal{J}$, as it is set to be $|\mathcal{B}|/2048$, and linearly warm it up to this value for 6 epochs followed by $90\%$ decay at 30, 60, and 80 epochs, training for 100 epochs total, as prescribed by \citet{buchlovsky2019tf}.
%(which is 0.125 in the case of $\mathcal{R}_{\text{WM}+\text{AD}}^{\text{TPE}}$)

\item Alignment learning rate $\eta \in [10^{-6}, 10^{-2}]$, sampled log-uniformly.
This parameter sets the adaptive learning rate on the Adam optimizer applied to the \emph{gradient} of the alignment loss $\mathcal{R}$, and which will be dropped synchronously by $90\%$ decay at 30, 60, and 80 epochs along with the Nesterov momentum learning rate on the \emph{pseudogradient} of the categorization objective $\mathcal{J}$.
% \textbf{$\mathcal{R}_{\text{WM}+\text{AD}}^{\text{TPE}}$ ended up setting $\mathbf{\eta = 0.005268466140227959}$}.

\item Number of delay epochs $de \in \{0, 1, 2\}$ for which we delay optimization of the categorization objective $\mathcal{J}$ and solely optimize the alignment loss $\mathcal{R}$.
If $de > 0$, we use the alignment learning rate $\eta$ during this delay period and the learning rate drops are shifted by $de$ epochs; otherwise, if $de = 0$, we linearly warmup $\eta$ for 6 epochs as well.
% \textbf{$\mathcal{R}_{\text{WM}+\text{AD}}^{\text{TPE}}$ ended up setting $\mathbf{de = 0}$ epochs.}

\item Whether or not to perform alternating minimization of $\mathcal{J}$ and $\mathcal{R}$ each step, or instead simultaneously optimize these objectives in a single training step. % \textbf{$\mathcal{R}_{\text{WM}+\text{AD}}^{\text{TPE}}$ ended up performing this alternation per step.}
\end{itemize}

The remaining metaparameters and their ranges were the same as those from Appendix \ref{sup:hp-wm-tpe-details}.

% \begin{itemize}
% \item $\mathcal{R}_{\text{WM}+\text{AD}}^{\text{TPE}}$ ended up setting the backward pass Gaussian input noise standard deviation $\mathbf{\sigma = 0.949983201505862}$.

% \item $\mathcal{R}_{\text{WM}+\text{AD}}^{\text{TPE}}$ ended up setting the ratio between the weighting of $\mathcal{P}^{\text{amp}}$ and $\mathcal{P}^{\text{decay}}$ given by $\mathbf{\alpha/\beta = 13.9035614049}$.

% \item $\mathcal{R}_{\text{WM}+\text{AD}}^{\text{TPE}}$ ended up setting the weighting of $\mathcal{P}^{\text{decay}}$ given by $\mathbf{\beta = 2.8109006032182933\times 10^{-8}}$.
% \end{itemize}

We fix the layer-wise operations as prescribed by \citet{akrout_deep_2019}, namely batch centering the backward path input and forward path outputs in the backward pass (\textbf{BI BWMC} and \textbf{FO BWMC}, respectively), as well as applying a ReLU activation function and bias to the forward path (\textbf{FO ReLU} and \textbf{FO Bias}, respectively) but not to the backward path in the backward pass (\textbf{BO ReLU} and \textbf{BO Bias}, respectively).

\subsubsection{$\mathcal{R}_{\text{WM}+\text{AD}+\text{OPS}}^{\text{TPE}}$ search space}
\label{sup:hp-wm-ad-ops-tpe-details}
Below is a description of the metaparameters and their ranges for the search that gave rise to $\mathcal{R}_{\text{WM}+\text{AD}+\text{OPS}}^{\text{TPE}}$ in Table~\ref{tab:hp-local}. 
In this search, we expand the search space described in Appendix \ref{sup:hp-wm-ad-tpe-details} to include boolean choices over layer-wise operations performed in the \emph{backward pass}, involving either the inputs, the forward path $f_l$ (involving only the forward weights $W_l$), or the backward path $b_l$ (involving only the backward weights $B_l$):

Use of biases in the forward and backward paths:
\begin{itemize}
\item \textbf{FO Bias:} Whether or not to use biases in the forward path.
%\textbf{$\mathcal{R}_{\text{WM}+\text{AD}+\text{OPS}}^{\text{TPE}}$ uses biases in the forward path.}

\item \textbf{BO Bias:} Whether or not to use biases in the backward path.
%\textbf{$\mathcal{R}_{\text{WM}+\text{AD}+\text{OPS}}^{\text{TPE}}$ does \emph{not} use biases in the backward path.}
\end{itemize}

Use of nonlinearities in the forward and backward paths:
\begin{itemize}
\item \textbf{FO ReLU:} Whether or not to apply a ReLU to the forward path output.
%\textbf{$\mathcal{R}_{\text{WM}+\text{AD}+\text{OPS}}^{\text{TPE}}$ applies a ReLU to the forward path output.}

\item \textbf{BO ReLU:} Whether or not to apply a ReLU to the backward path output.
%\textbf{$\mathcal{R}_{\text{WM}+\text{AD}+\text{OPS}}^{\text{TPE}}$ applies a ReLU to the backward path output.}
\end{itemize}

Centering and normalization operations in the forward and backward paths:
\begin{itemize}
\item \textbf{FO BWMC:} Whether or not to mean center (across the \emph{batch} dimension) the forward path output $f_l = f_l - \bar{f}_l$.
%\textbf{$\mathcal{R}_{\text{WM}+\text{AD}+\text{OPS}}^{\text{TPE}}$ applies batch-wise mean centering to the forward path output.}

\item \textbf{BI BWMC:} Whether or not to mean center (across the \emph{batch} dimension) the backward path \emph{input}. 
%\textbf{$\mathcal{R}_{\text{WM}+\text{AD}+\text{OPS}}^{\text{TPE}}$ applies batch-wise mean centering to the backward path input.}

\item \textbf{FO FWMC:} Whether or not to mean center (across the \emph{feature} dimension) the forward path output $f_l = f_l - \hat{f}_l$.
%\textbf{$\mathcal{R}_{\text{WM}+\text{AD}+\text{OPS}}^{\text{TPE}}$ applies feature-wise mean centering to the forward path output.}

\item \textbf{BO FWMC:} Whether or not to mean center (across the \emph{feature} dimension) the backward path output $b_l = b_l - \hat{b}_l$.
%\textbf{$\mathcal{R}_{\text{WM}+\text{AD}+\text{OPS}}^{\text{TPE}}$ does \emph{not} apply feature-wise mean centering to the backward path output.}

\item \textbf{FO FWL2N:} Whether or not to L2 normalize (across the feature dimension) the forward path output $f_l = (f_l - \hat{f}_l)/||f_l - \hat{f}_l||_2$. 
%\textbf{$\mathcal{R}_{\text{WM}+\text{AD}+\text{OPS}}^{\text{TPE}}$ does \emph{not} apply feature-wise L2 normalization to the forward path output.}

\item \textbf{BO FWL2N:} Whether or not to L2 normalize (across the feature dimension) the backward path output $b_l = (b_l - \hat{b}_l)/||b_l - \hat{b}_l||_2$. 
%\textbf{$\mathcal{R}_{\text{WM}+\text{AD}+\text{OPS}}^{\text{TPE}}$ applies feature-wise L2 normalization to the backward path output.}
\end{itemize}

Centering and normalization operations applied to the inputs to the backward pass:
\begin{itemize}
\item \textbf{BI FWMC:} Whether or not to mean center (across the feature dimension) the backward pass input $x_l = x_l - \hat{x}_l$.
%\textbf{$\mathcal{R}_{\text{WM}+\text{AD}+\text{OPS}}^{\text{TPE}}$ does \emph{not} apply feature-wise mean centering to the backward pass input.}

\item \textbf{BI FWL2N:} Whether or not to L2 normalize (across the feature dimension) the backward pass input $x_l = (x_l - \hat{x}_l)/||x_l - \hat{x}_l||_2$. 
%\textbf{$\mathcal{R}_{\text{WM}+\text{AD}+\text{OPS}}^{\text{TPE}}$ does \emph{not} apply feature-wise L2 normalization to the backward pass input.}
\end{itemize}

The remaining metaparameters and their ranges were the same as those from Appendix \ref{sup:hp-wm-ad-tpe-details}.
% \begin{itemize}
% \item $\mathcal{R}_{\text{WM}+\text{AD}+\text{OPS}}^{\text{TPE}}$ ended up setting the training batch size $\mathbf{|\mathcal{B}| = 256}$. Therefore, as explained in Appendix \ref{sup:hp-wm-ad-tpe-details}, the forward path Nesterov momentum learning rate is 0.125.

% \item $\mathcal{R}_{\text{WM}+\text{AD}+\text{OPS}}^{\text{TPE}}$ ended up setting the alignment Adam learning rate to $\mathbf{\eta = 0.0024814785849458275}$.

% \item $\mathcal{R}_{\text{WM}+\text{AD}+\text{OPS}}^{\text{TPE}}$ ended up setting the number of delay epochs to be $\mathbf{de = 0}$ epochs.

% \item $\mathcal{R}_{\text{WM}+\text{AD}+\text{OPS}}^{\text{TPE}}$ ended up performing alternating minimization of $\mathcal{J}$ and $\mathcal{R}$ per training step.

% \item $\mathcal{R}_{\text{WM}+\text{AD}+\text{OPS}}^{\text{TPE}}$ ended up setting the backward pass Gaussian input noise standard deviation $\mathbf{\sigma = 0.6401550883361206}$.

% \item $\mathcal{R}_{\text{WM}+\text{AD}+\text{OPS}}^{\text{TPE}}$ ended up setting the ratio between the weighting of $\mathcal{P}^{\text{amp}}$ and $\mathcal{P}^{\text{decay}}$ given by $\mathbf{\alpha/\beta = 0.13444256251}$.

% \item $\mathcal{R}_{\text{WM}+\text{AD}+\text{OPS}}^{\text{TPE}}$ ended up setting the weighting of $\mathcal{P}^{\text{decay}}$ given by $\mathbf{\beta = 232.78563021186207}$.
% \end{itemize}

\subsubsection{$\mathcal{R}_{\text{IA}}^{\text{TPE}}$ search space}
\label{sup:hp-ia-tpe-details}
Below is a description of the metaparameters and their ranges for the search that gave rise to $\mathcal{R}_{\text{IA}}^{\text{TPE}}$ in Table~\ref{tab:hp-local}. In this search, we expand the search space described in Appendix \ref{sup:hp-wm-ad-ops-tpe-details}, to now include the additional $\mathcal{P}^{\text{null}}$ primitive.

\begin{itemize}
\item The weighting of $\mathcal{P}^{\text{null}}$ given by $\gamma \in [10^{-11}, 10^{7}]$, sampled log-uniformly.
%\textbf{$\mathcal{R}_{\text{IA}}^{\text{TPE}}$ ended up setting $\mathbf{\gamma = 3.1610192645608835\times 10^{-6}}$.}
\end{itemize}

The remaining metaparameters and their ranges were the same as those from Appendix \ref{sup:hp-wm-ad-ops-tpe-details}.

\subsection{Symmetric and Activation Alignment metaparameters}
\label{sup:non-local-hp}
We now describe the metaparameters used to generate Table~\ref{tab:empimnet}. 
We used a batch size of 256, forward path Nesterov with Momentum of 0.9 and a learning rate of 0.1 applied to the categorization objective $\mathcal{J}$, warmed up linearly for 5 epochs, with learning rate drops at 30, 60, and 80 epochs, trained for a total of 90 epochs, as prescribed by \citet{he_deep_2016}.

For Symmetric and Activation Alignment ($\mathcal{R}_{\text{SA}}$ and $\mathcal{R}_{\text{AA}}$), we used Adam on the alignment loss $\mathcal{R}$ with a learning rate of $0.001$, along with the following weightings for their primitives:
\begin{itemize}
\item Symmetric Alignment: $\alpha=10^{-3}, \beta=2\times 10^{-3}$
\item Activation Alignment: $\alpha=10^{-3}, \beta=2\times 10^{-3}$
\end{itemize}
We use biases in both the forward and backward paths of the backward pass, but do \emph{not} employ a ReLU nonlinearity to either path.

\subsection{Noisy updates}
\label{sup:noisy-updates}

We describe the experimental setup and metaparameters used in \S\ref{sec:non-local-learning-rules} to generate Fig.~\ref{fig:noise-plot}.  

Fig.~\ref{fig:noise-plot-a} was generated by running 10 trials for each experiment configuration. 
The error bars show the standard error of the mean across trials. 

For backpropagation  we used a momentum optimizer with an initial learning rate of $0.1$, standard batch size of $256$, and learning rate drops at $30$ and $60$ epochs.

For Symmetric and Activation Alignment we used the same metaparameters as backpropagation for the categorization objective $\mathcal{J}$ and an Adam optimizer with an initial learning rate of $0.001$ and learning rate drops at $30$ and $60$ epochs for the alignment loss $\mathcal{R}$.  All other metaparameters were the same as described in Appendix \ref{sup:non-local-hp}.

\begin{figure}[tb]
\vskip 0.2in
\centering
\includegraphics[width=0.48\textwidth]{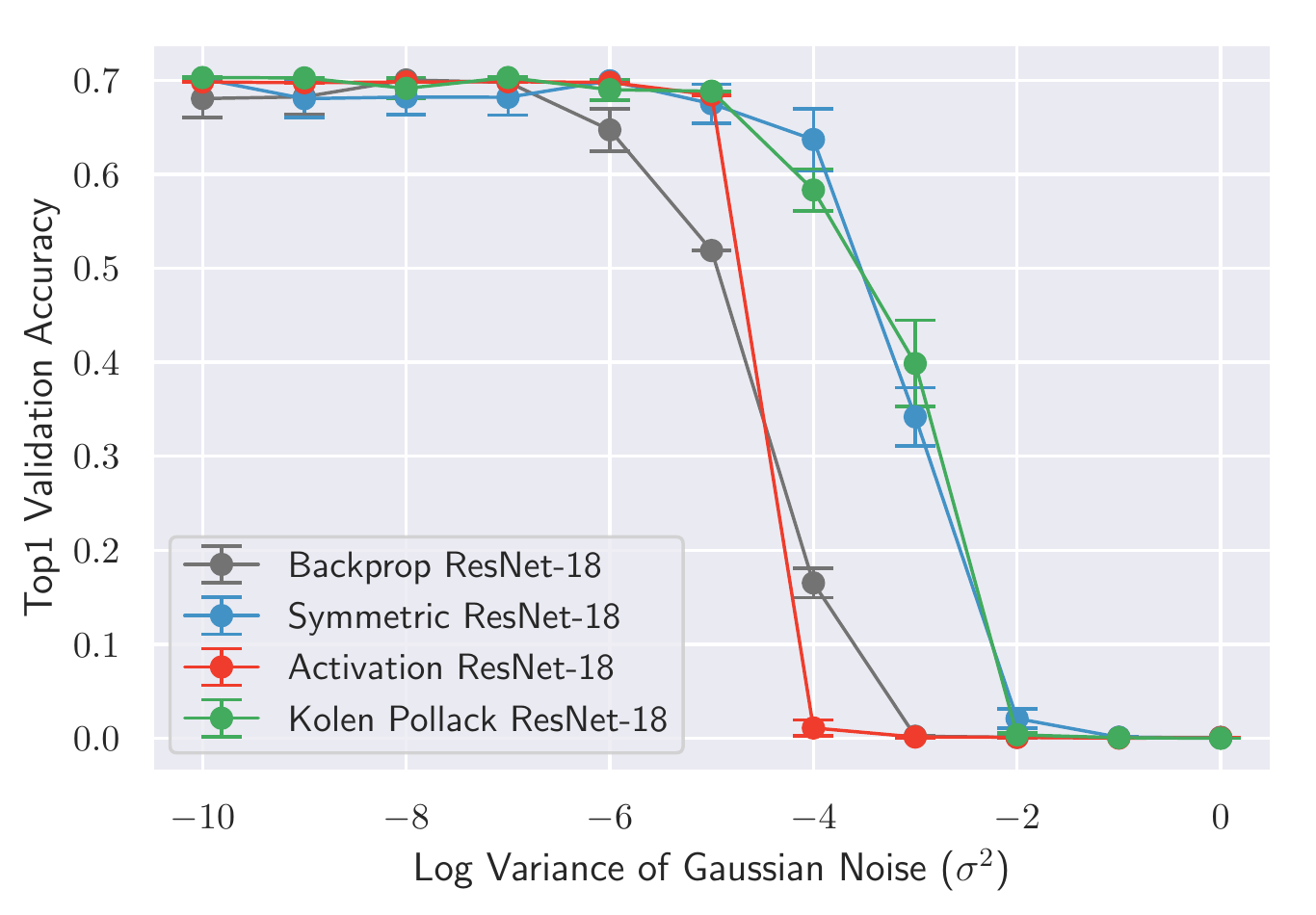}
\caption{\textbf{Noisy updates.} Symmetric Alignment, Activation Alignment, and Kolen-Pollack are \emph{more} robust to noisy updates than backpropagation for ResNet-18.}
\label{fig:all-noise}
\vskip -0.2in
\end{figure}

In all experiments we added the noise to the update given by the respective optimizers and scaled by the current learning rate, that way at learning rate drops the noise scaled appropriately.  To account for the fact that the initial learning rate for the backpopagation experiments was $0.1$, while for symmetric and activation experiments it was $0.001$, we shifted the latter two curves by $10^4$ to account for the effective difference in variance.
See Fig.~\ref{fig:all-noise}.

\subsection{Metaparameter importance quantification}
\label{sup:meta-analysis-hp}
\begin{figure}
\centering
\includegraphics[width=0.96\columnwidth]{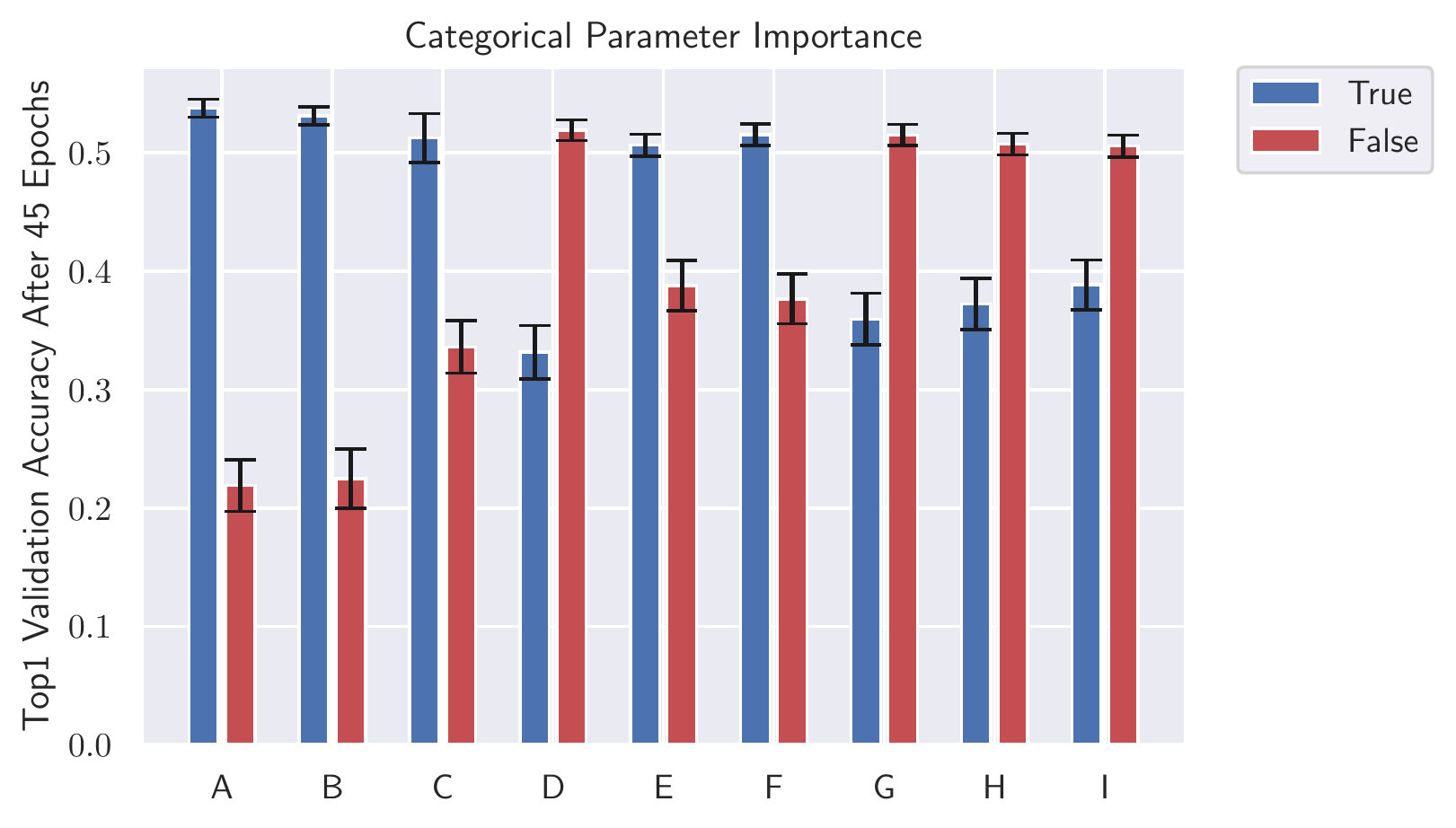}
\caption{\textbf{Analysis of important categorical metaparameters of top performing local rule $\mathcal{R}_{\text{IA}}^{\text{TPE}}$.} Mean across models, and the error bars indicate SEM across models.}
\label{fig:hp-metaanalysis}
\end{figure}

We include here the set of discrete metaparameters that mattered the most across hundreds of models in our large-scale search, sorted by most to least important, plotted in Fig.~\ref{fig:hp-metaanalysis}. 
Specifically, these amount to choices of activation, layer-wise normalization, input normalization, and Gaussian noise in the forward and backward paths of the backward pass. 
The detailed labeling is given as follows: \textbf{A:} Whether or not to L2 normalize (across the feature dimension) the backward path outputs in the backward pass. 
\textbf{B:} Whether to use Gaussian noise in the backward pass inputs. 
\textbf{C:} Whether to solely optimize the alignment loss in the first 1-2 epochs of training. 
\textbf{D, E:} Whether or not to apply a non-linearity in the backward or forward path outputs in the backward pass, respectively. 
\textbf{F:} Whether or not to apply a bias in the forward path outputs (pre-nonlinearity). 
\textbf{G, H:} Whether or not to mean center or L2 normalize (across the feature dimension) the inputs to the backward pass. 
\textbf{I:} Same as \textbf{A}, but instead applied to the forward path outputs in the backward pass.

\subsection{Neural Fitting Procedure}
\label{sup:neural-fit}
% In Fig.~\ref{fig:neural-fits}, ``FA'' refers to feedback alignment \cite{lillicrap_random_2016}. ``Mirror'' refers to $\mathcal{R}_{\text{WM}}$ using the \cite{akrout_deep_2019} metaparameters in \S\ref{sec:local-learning-rules}. ``Information'' refers to $\mathcal{R}_{\text{IA}}^{\text{TPE}}$ obtained through TPE in \S\ref{sec:local-learning-rules}. ``Symmetric'' and ``Activation'' correspond to $\mathcal{R}_{\text{SA}}$ and $\mathcal{R}_{\text{AA}}$ in \S\ref{sec:non-local-learning-rules}.

We fit trained model features to multi-unit array responses from \cite{majaj2015simple}.
Briefly, we fit to 256 recorded sites from two monkeys. These came from three multi-unit arrays per monkey: one implanted in V4, one in posterior IT, and one in central and anterior IT.
Each image was presented approximately 50 times, using rapid visual stimulus presentation (RSVP).
Each stimulus was presented for 100 ms, followed by a mean gray background interleaved between images.
Each trial lasted 250 ms. The image set consisted of 5120 images based on 64 object categories.
Each image consisted of a 2D projection of a 3D model added to a random background.
The pose, size, and $x$- and $y$-position of the object was varied across the image set, whereby 2 levels of variation were used (corresponding to medium and high variation from \cite{majaj2015simple}.)
Multi-unit responses to these images were binned in 10ms windows, averaged across trials of the same image, and normalized to the average response to a blank image.
They were then averaged 70-170 ms post-stimulus onset, producing a set of (5120 images x 256 units) responses, which were the targets for our model features to predict.
The 5120 images were split 75-25 within each object category into a training set and a held-out testing set.

\section{Visualizations}
\label{sup:weight-scatter}

\begin{figure}[t]%[H] %idk why this option makes all figures appear at the end of the doc...
\vskip 0.2in
% Symmetric
\begin{subfigure}{0.3\columnwidth}
    \centering
    \includegraphics[width=\textwidth]{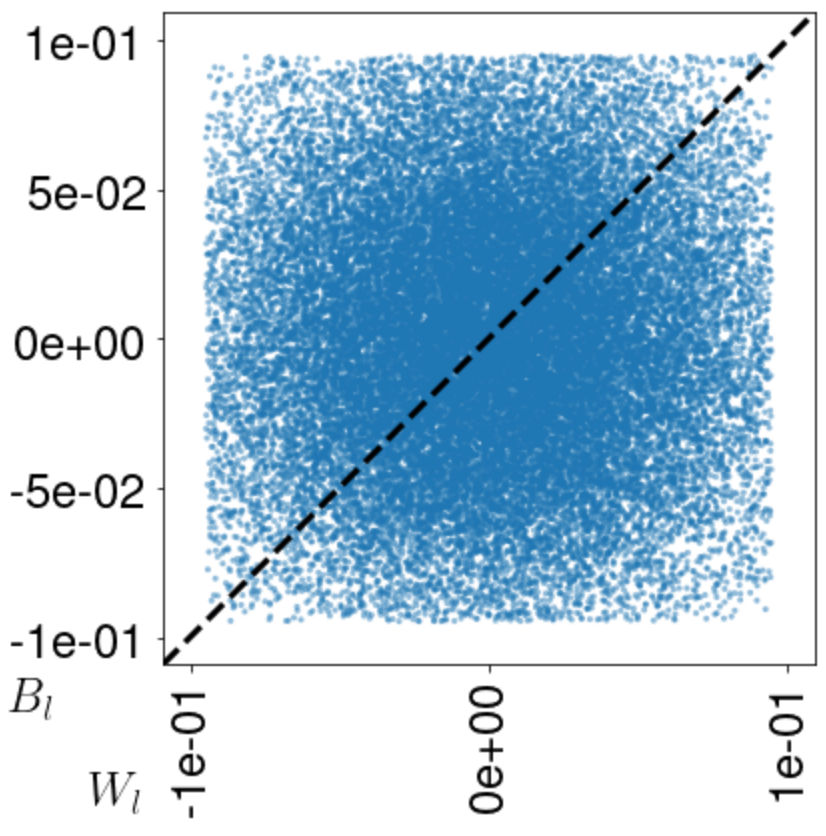}
    \caption{SA Epoch 0}
\end{subfigure}
\begin{subfigure}{0.3\columnwidth}
    \centering
    \includegraphics[width=\textwidth]{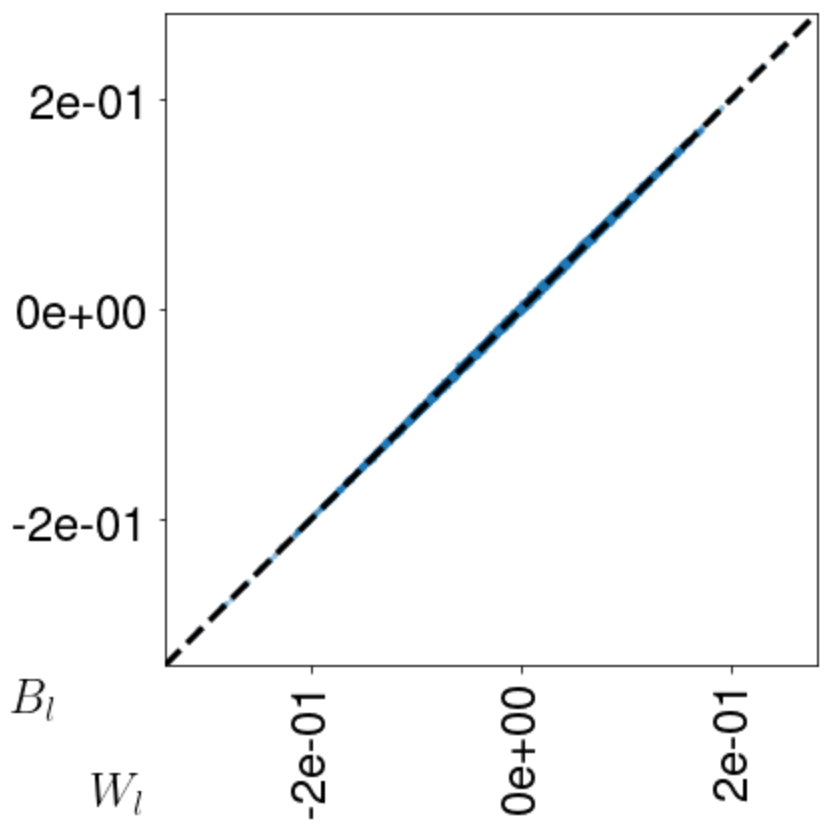}
    \caption{SA Epoch 2}
\end{subfigure}
\begin{subfigure}{0.3\columnwidth}
    \centering
    \includegraphics[width=\textwidth]{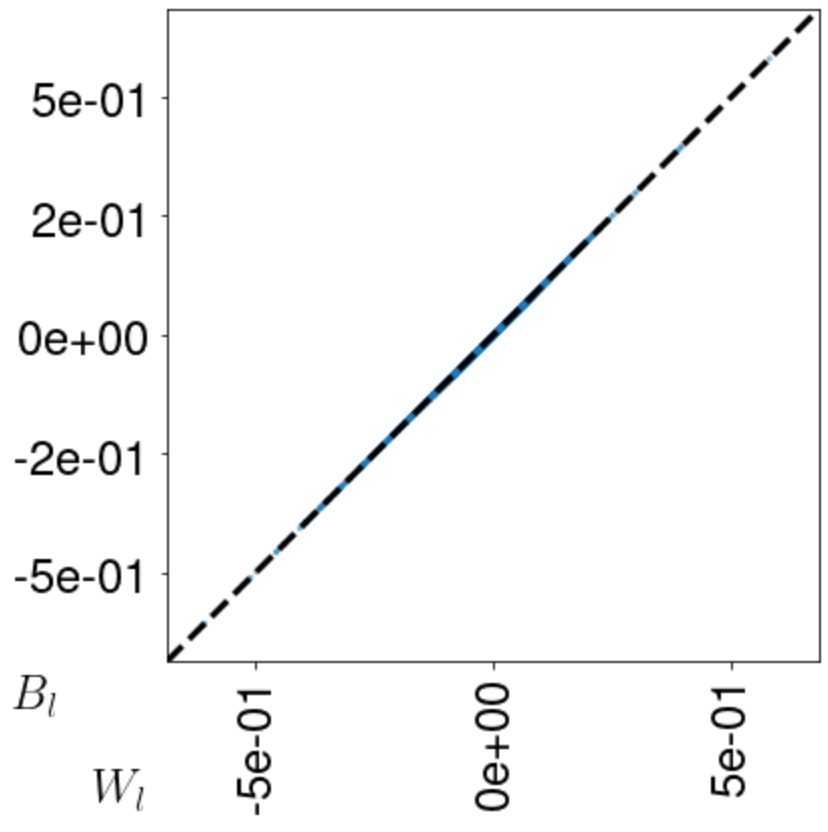}
    \caption{SA Epoch 90}
\end{subfigure}

% Activation
\begin{subfigure}{0.3\columnwidth}
    \centering
    \includegraphics[width=\textwidth]{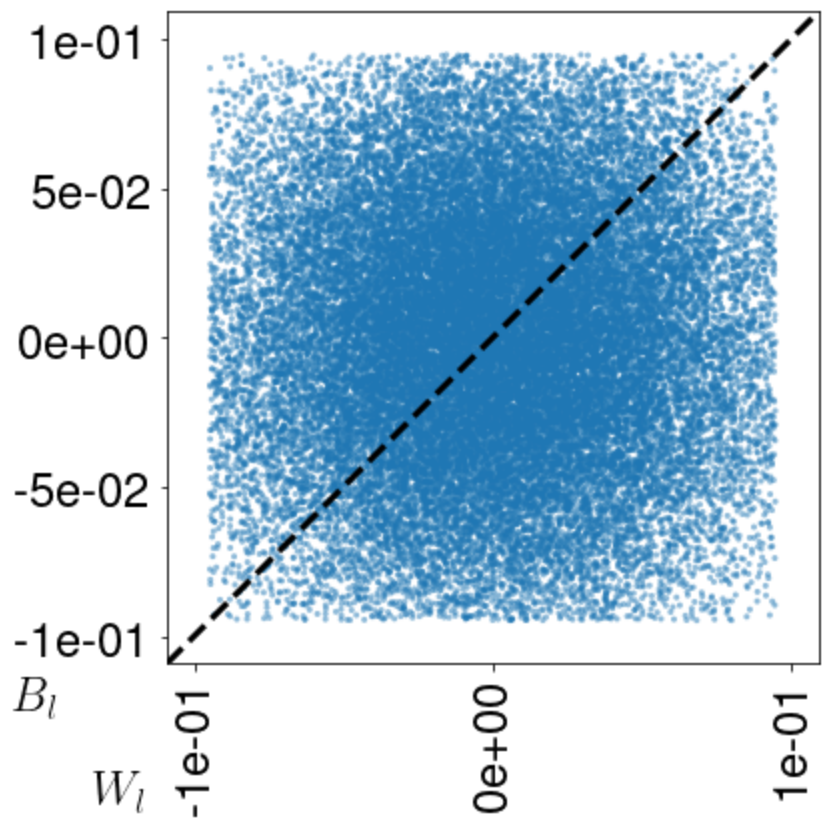}
    \caption{AA Epoch 0}
\end{subfigure}
\begin{subfigure}{0.3\columnwidth}
    \centering
    \includegraphics[width=\textwidth]{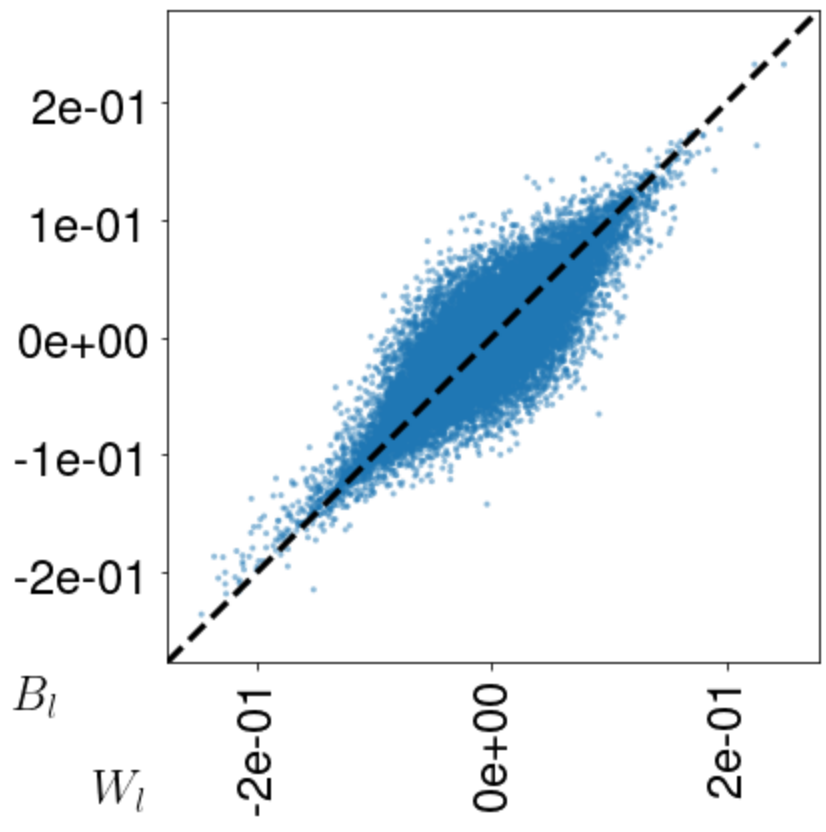}
    \caption{AA Epoch 2}
\end{subfigure}
\begin{subfigure}{0.3\columnwidth}
    \centering
    \includegraphics[width=\textwidth]{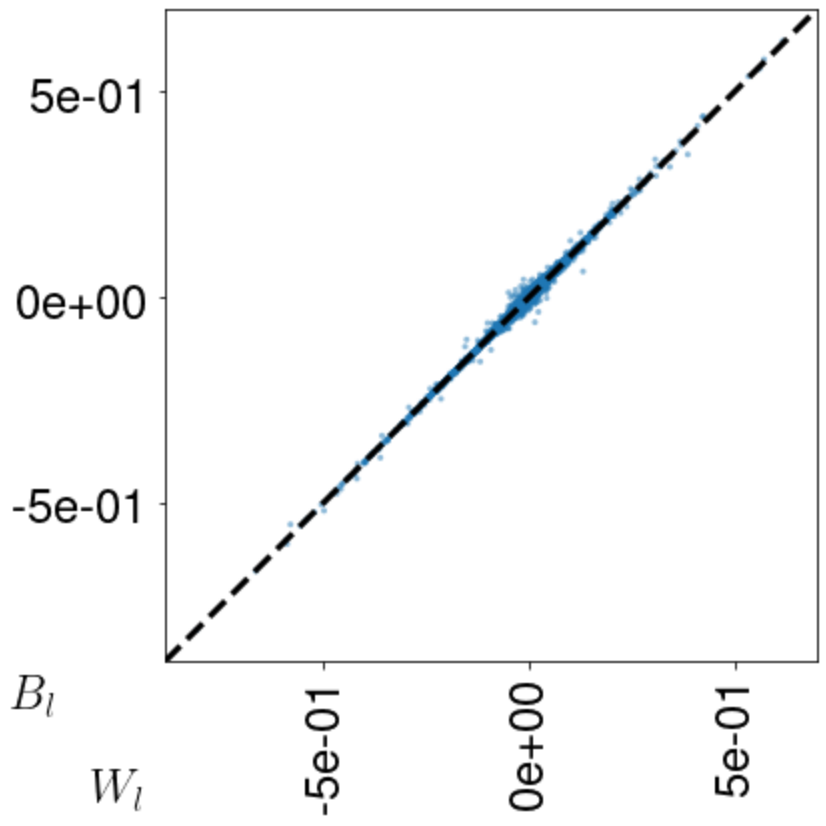}
    \caption{AA Epoch 90}
\end{subfigure}

% Vanilla Weight Mirror
\begin{subfigure}{0.3\columnwidth}
    \centering
    \includegraphics[width=\textwidth]{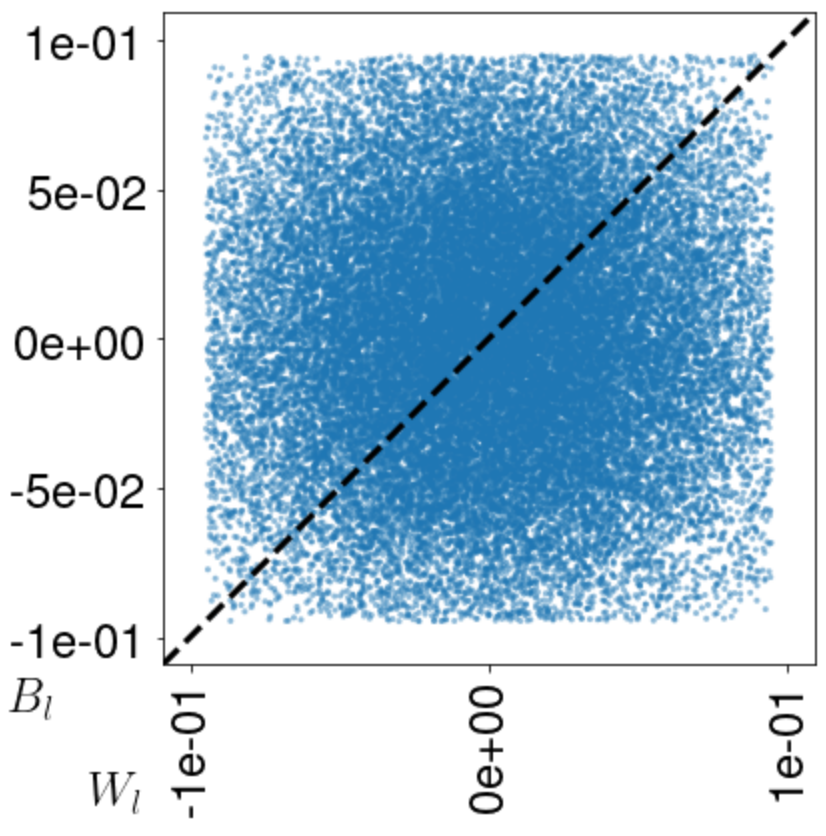}
    \caption{WM Epoch 0}
\end{subfigure}
\begin{subfigure}{0.3\columnwidth}
    \centering
    \includegraphics[width=\textwidth]{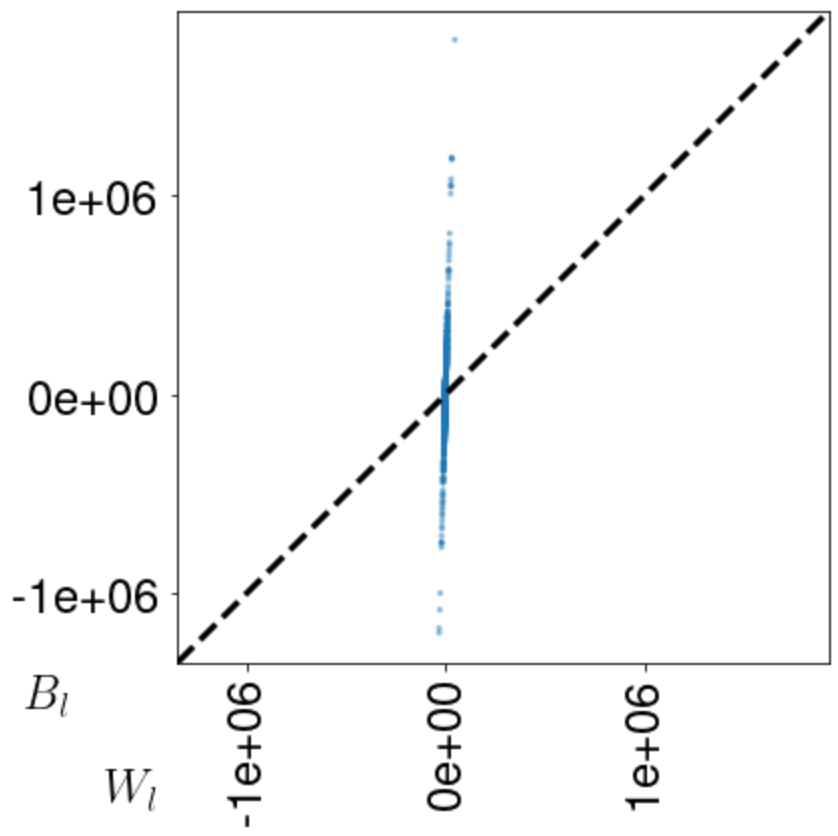}
    \caption{WM Epoch 2}
\end{subfigure}
\begin{subfigure}{0.3\columnwidth}
    \centering
    \includegraphics[width=\textwidth]{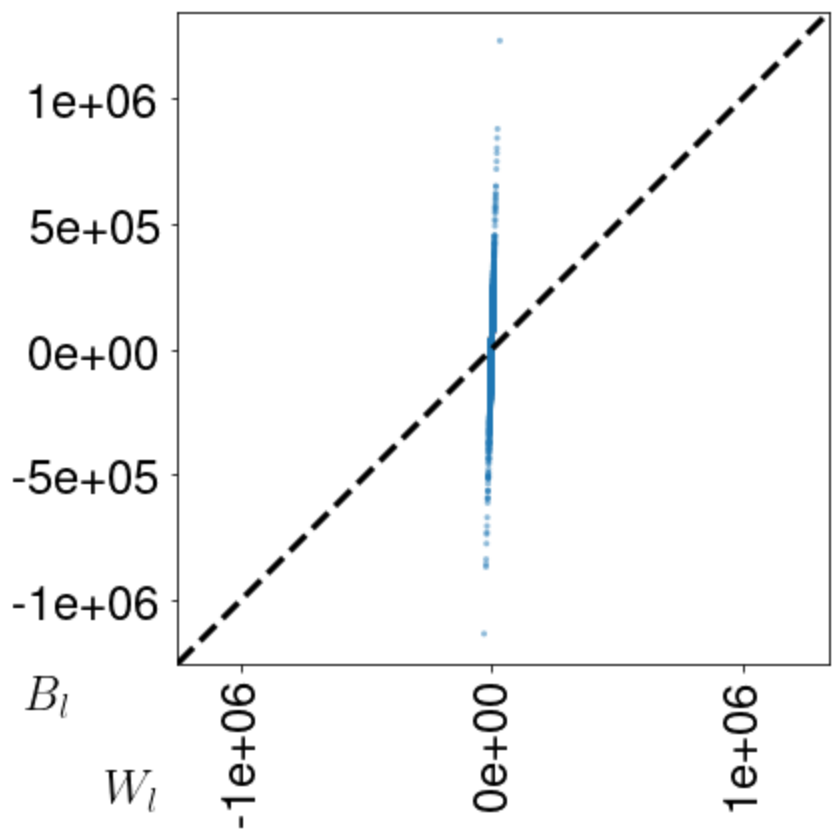}
    \caption{WM Epoch 90}
\end{subfigure}

% Best TPE Information
\begin{subfigure}{0.3\columnwidth}
    \centering
    \includegraphics[width=\textwidth]{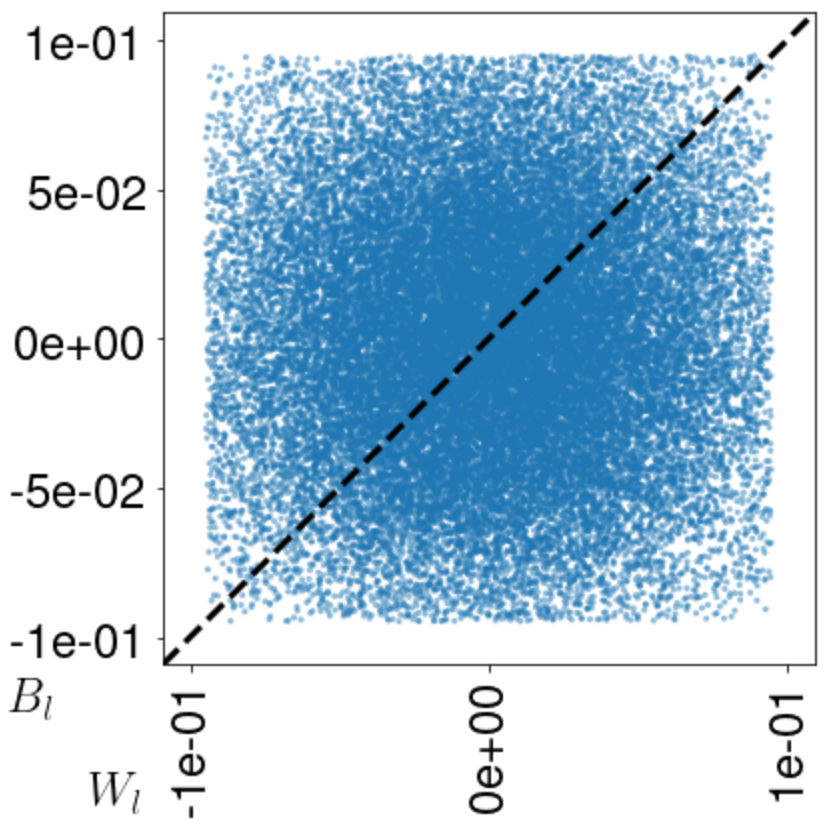}
    \caption{IA Epoch 0}
\end{subfigure}
\begin{subfigure}{0.3\columnwidth}
    \centering
    \includegraphics[width=\textwidth]{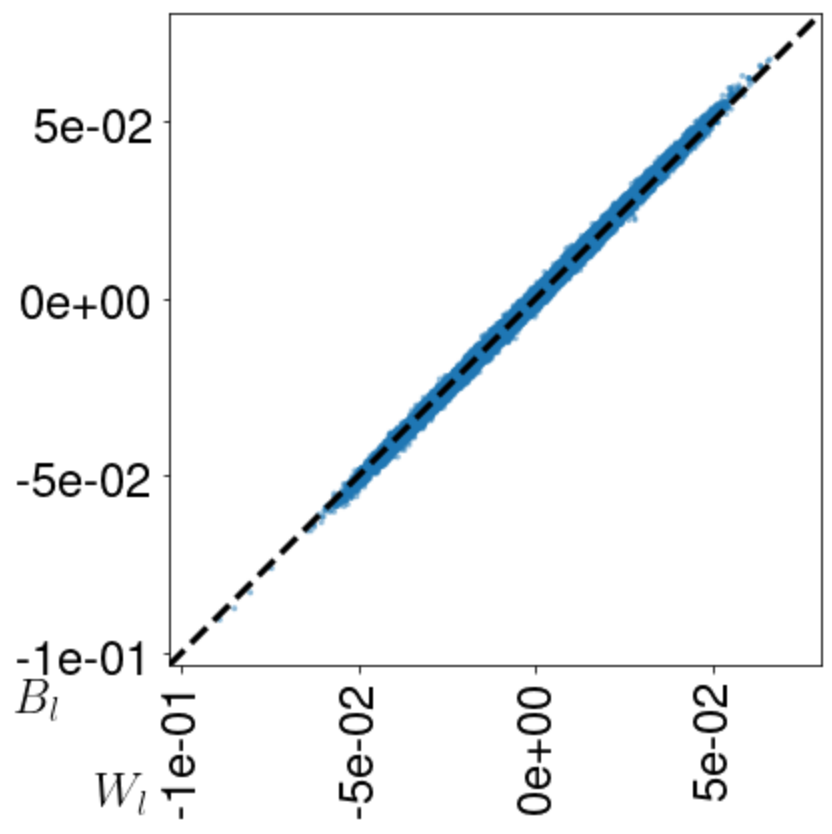}
    \caption{IA Epoch 2}
\end{subfigure}
\begin{subfigure}{0.3\columnwidth}
    \centering
    \includegraphics[width=\textwidth]{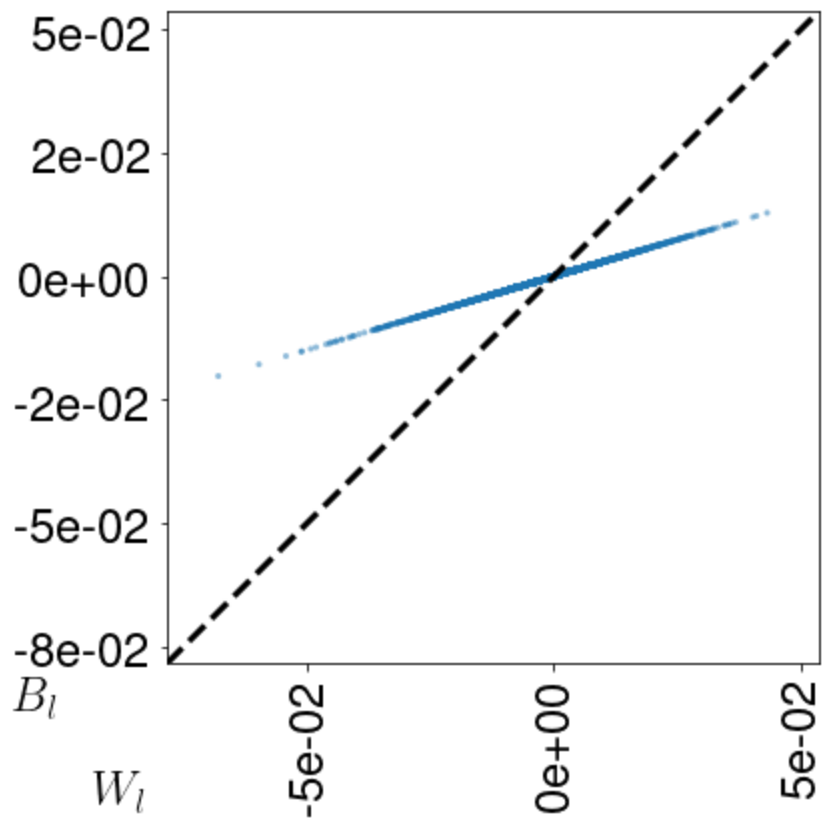}
    \caption{IA Epoch 90}
\end{subfigure}
\caption{\textbf{Learning symmetry.} Weight values of the third convolutional layer in ResNet-18 throughout training with various learning rules.
Each dot represents an element in layer $l$'s weight matrix and its $(x,y)$ location corresponds to its forward and backward weight values, $(W_l^{(i,j)},B_l^{(j,i)} )$. The dotted diagonal line shows perfect weight symmetry, as is the case in backpropagation.
\label{fig:full_weight_scatter}}
\end{figure}

\begin{figure*}
\begin{subfigure}{2\columnwidth}
    \centering
    \includegraphics[width=0.46\textwidth]{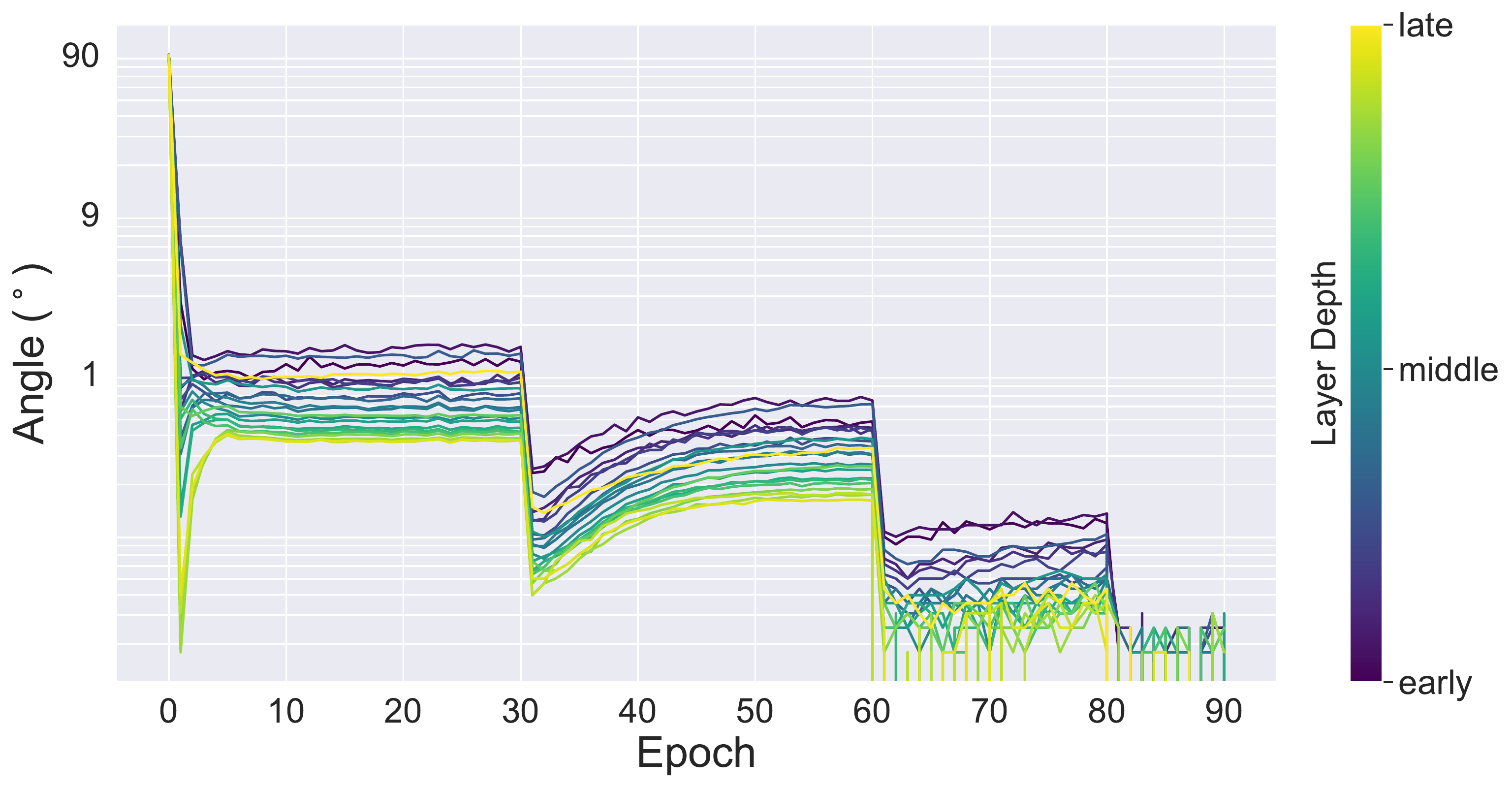}
    \includegraphics[width=0.49\textwidth]{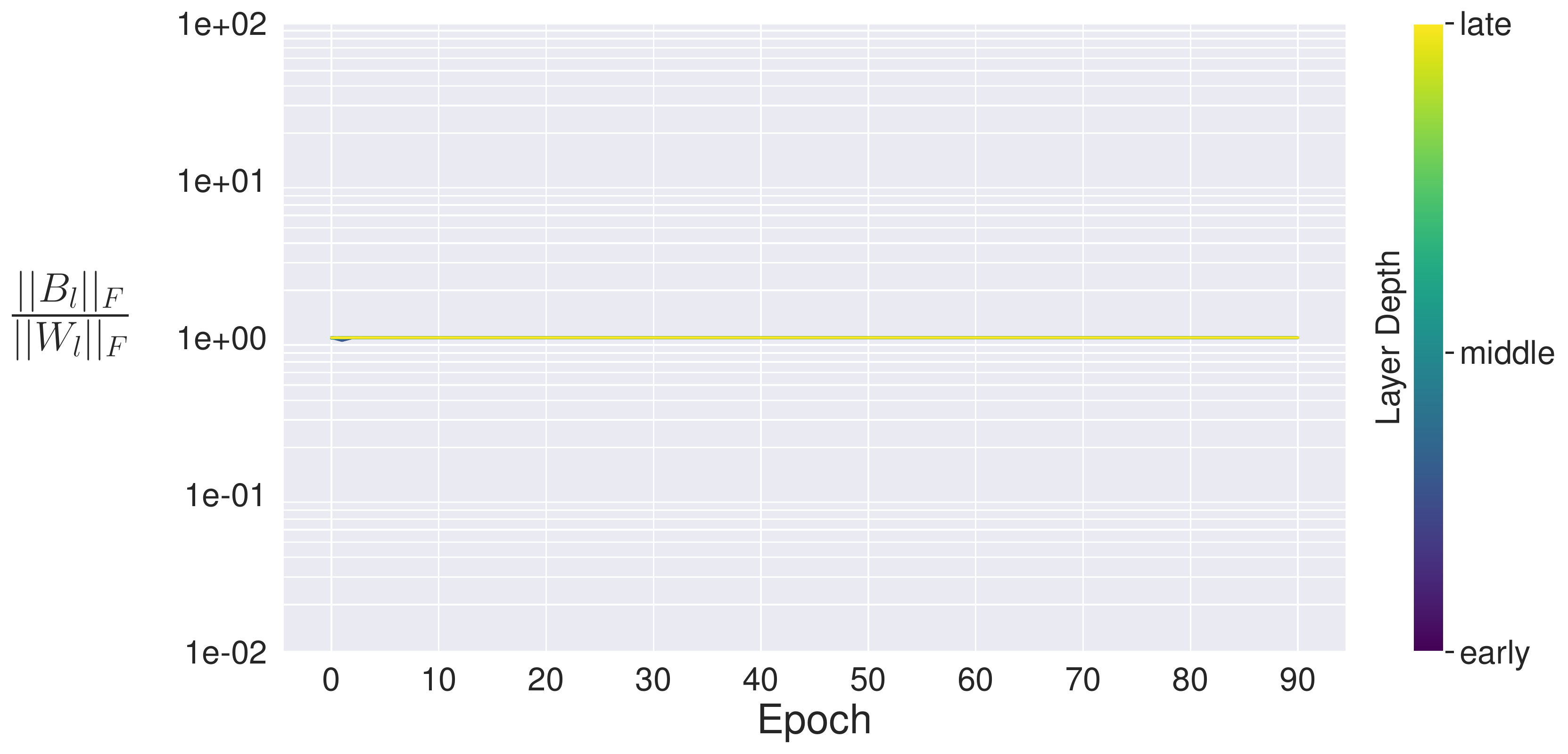}
    \caption{Symmetric Alignment}
\end{subfigure}
\begin{subfigure}{2\columnwidth}
    \centering
    \includegraphics[width=0.46\textwidth]{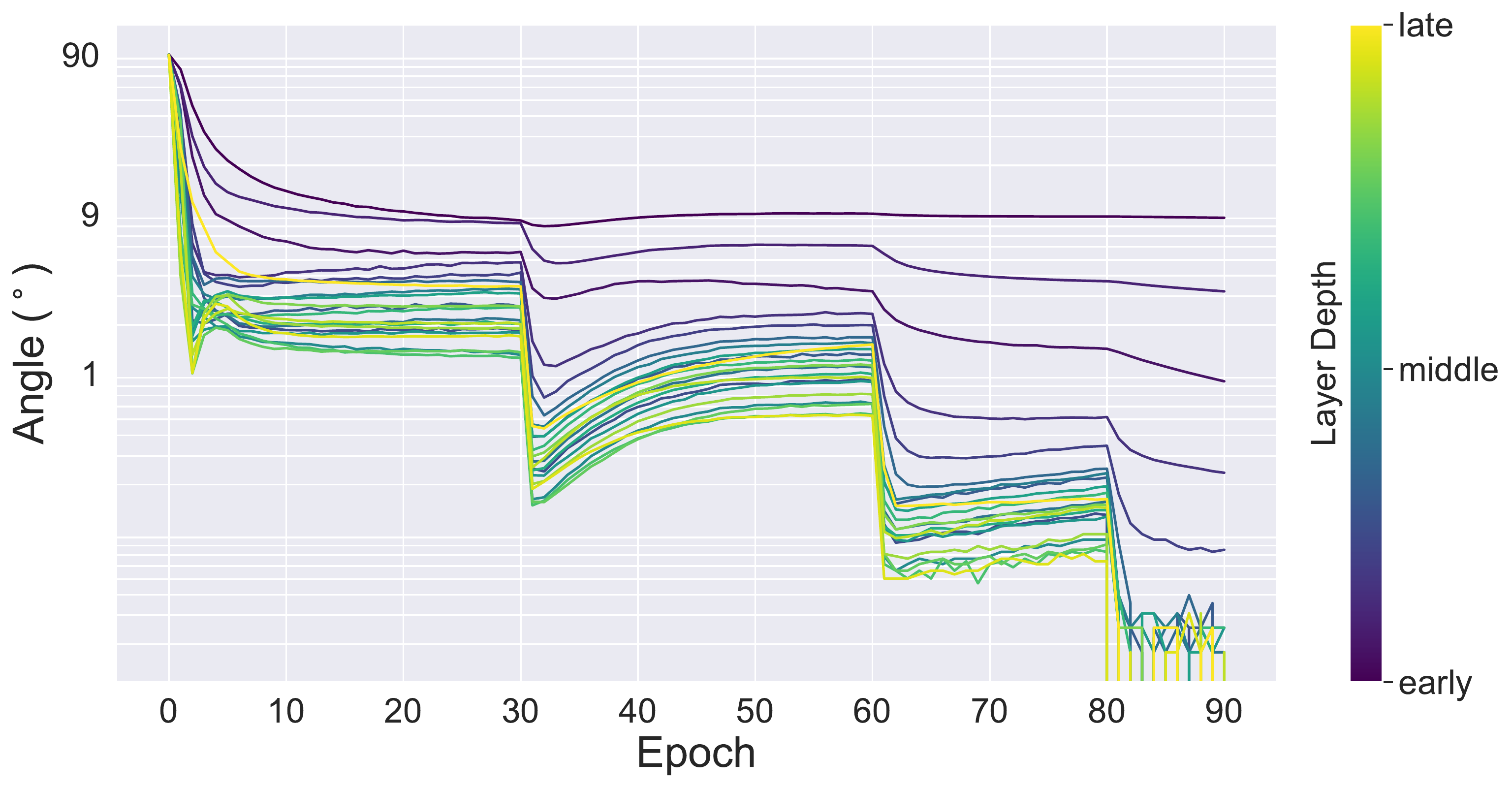}
    \includegraphics[width=0.49\textwidth]{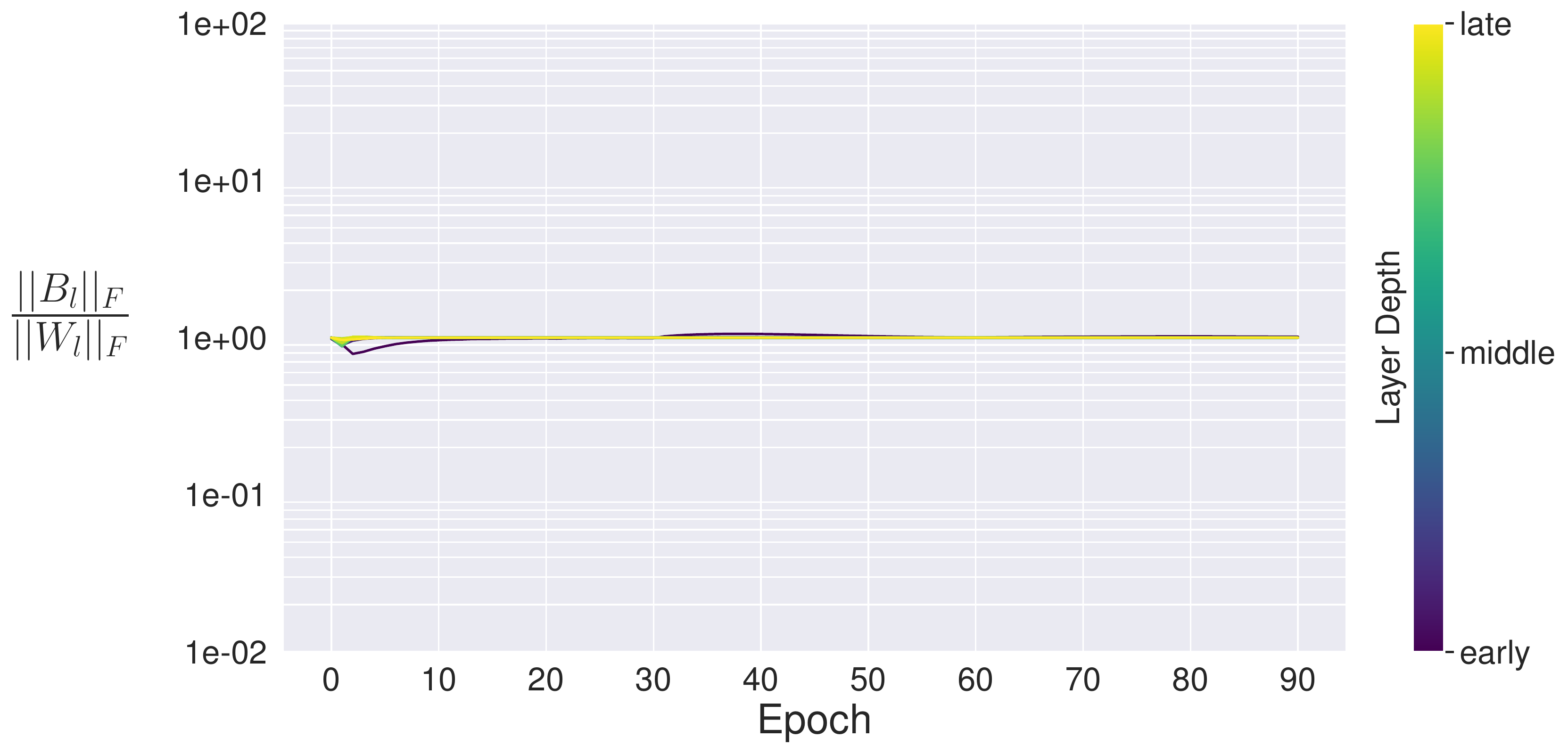}
    \caption{Activation Alignment}
\end{subfigure}
\begin{subfigure}{2\columnwidth}
    \centering
    \includegraphics[width=0.46\textwidth]{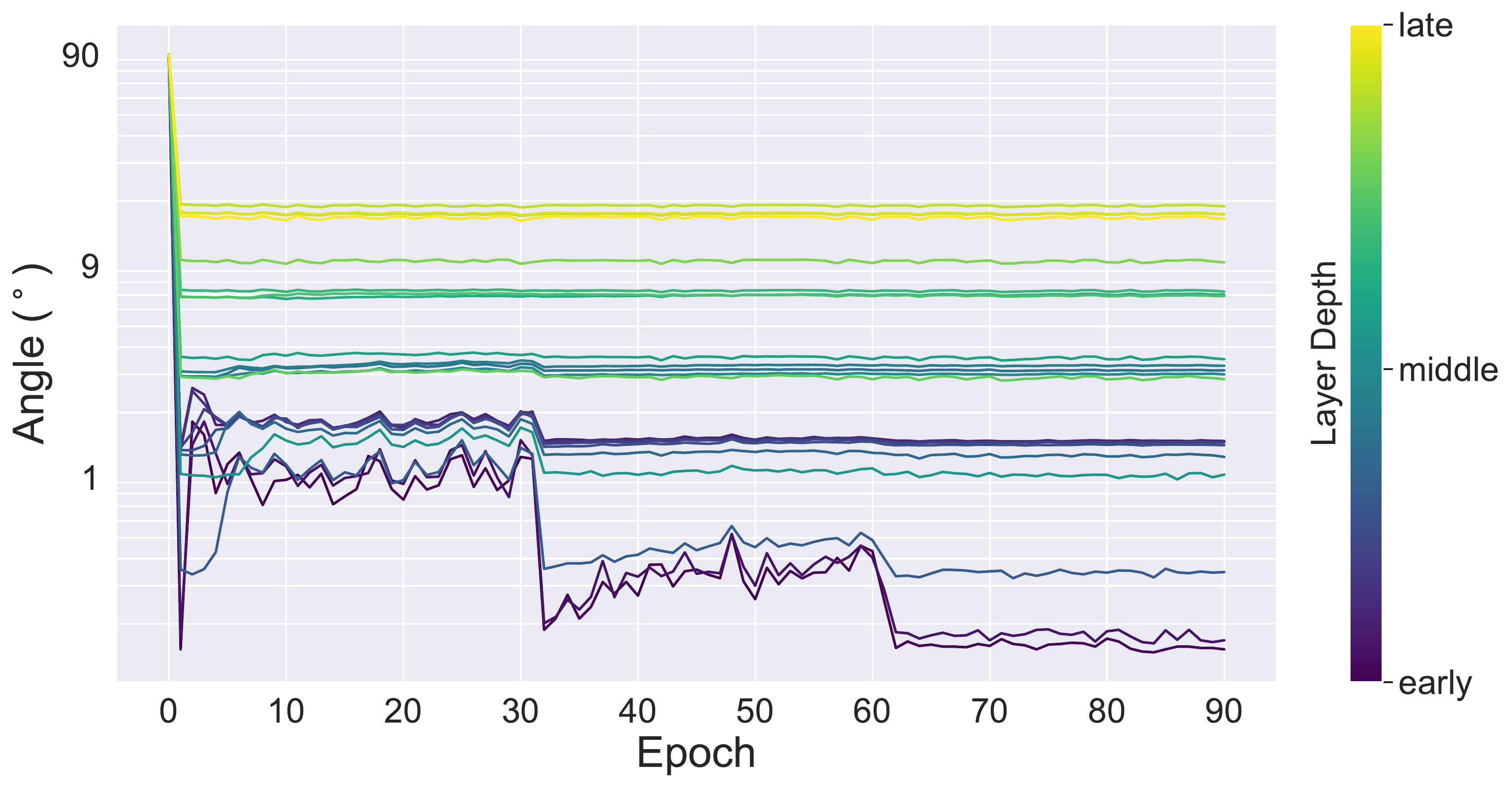}
    \includegraphics[width=0.49\textwidth]{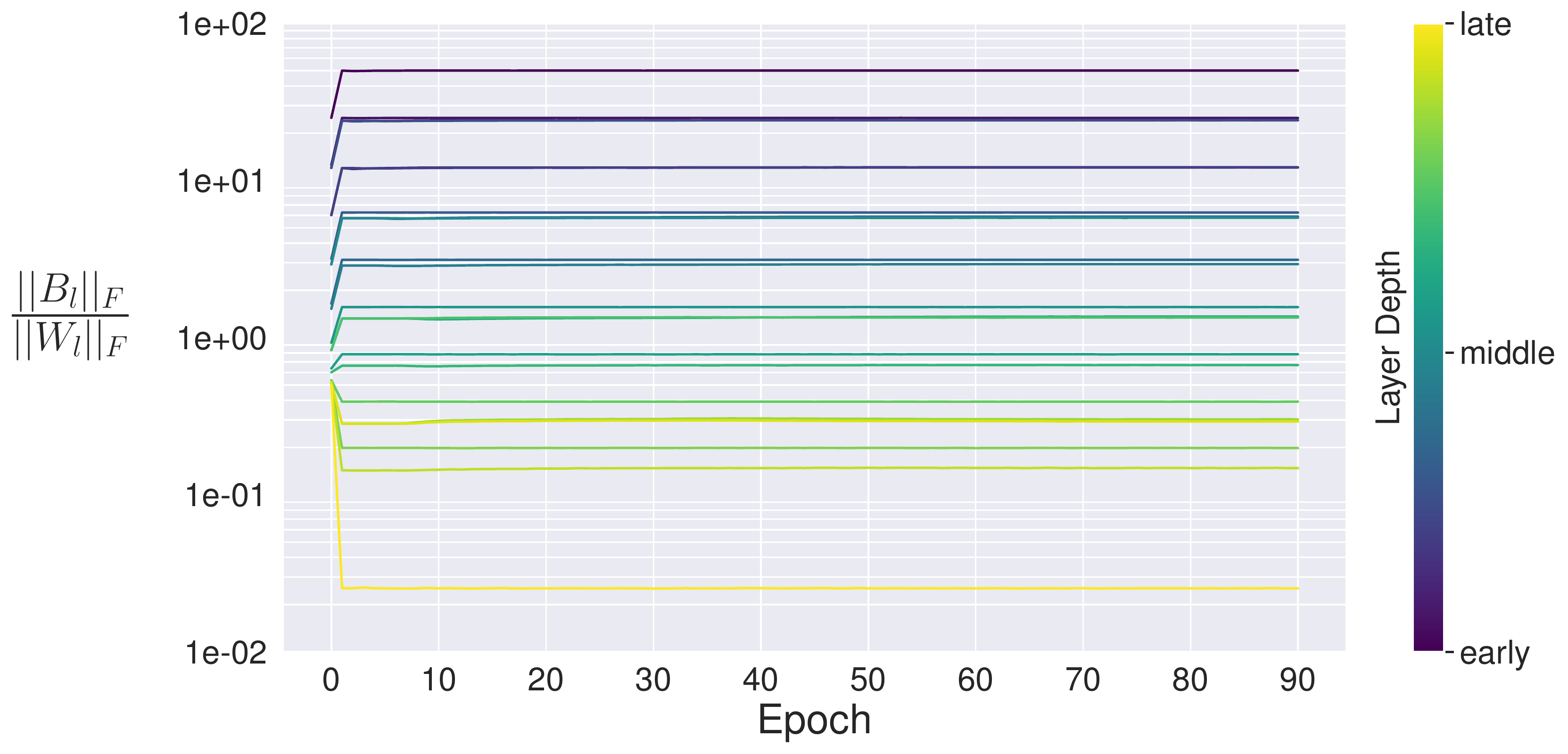}
    \caption{Weight Mirror}
\end{subfigure}
\begin{subfigure}{2\columnwidth}
    \centering
    \includegraphics[width=0.46\textwidth]{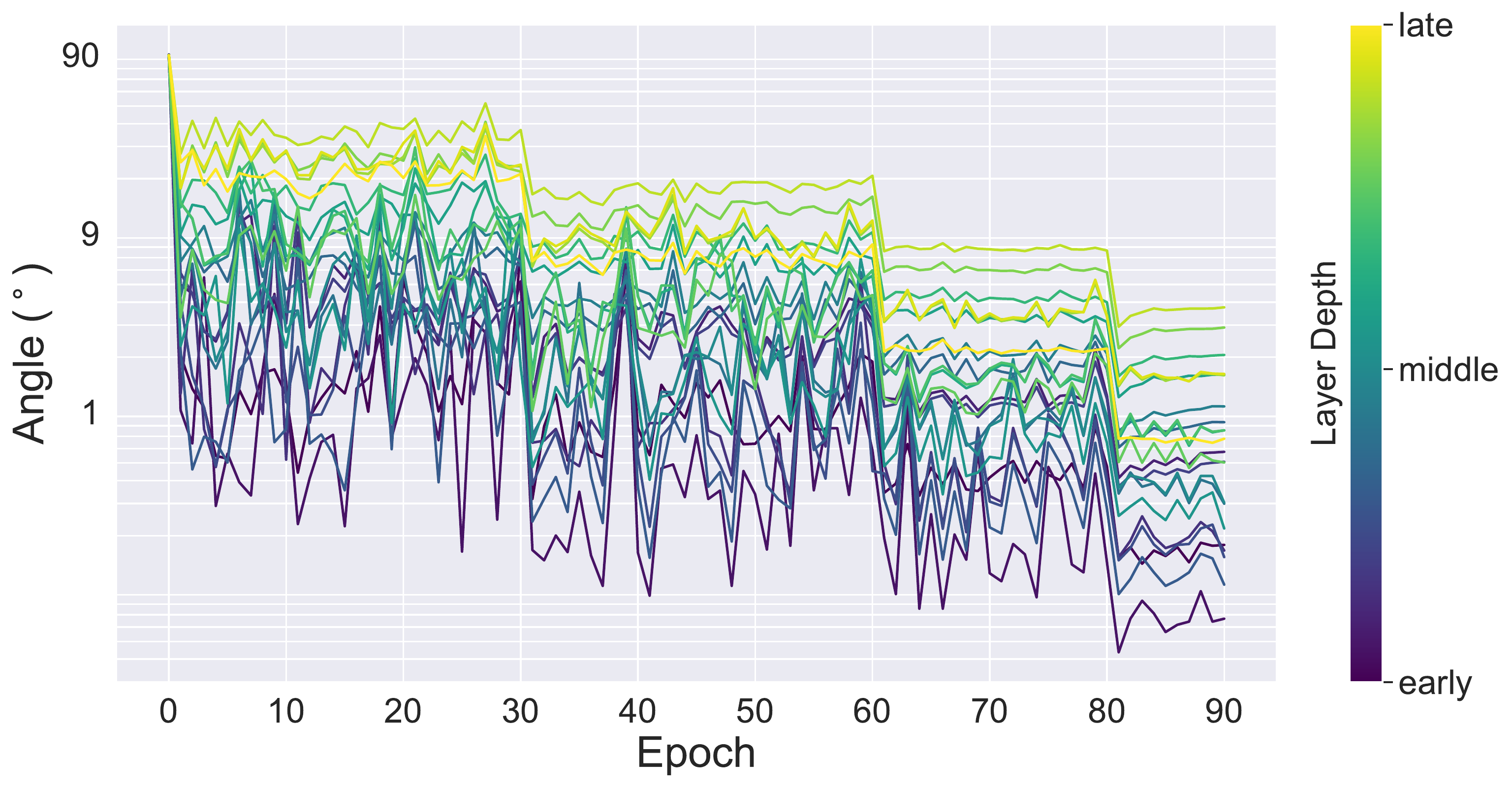}
    \includegraphics[width=0.49\textwidth]{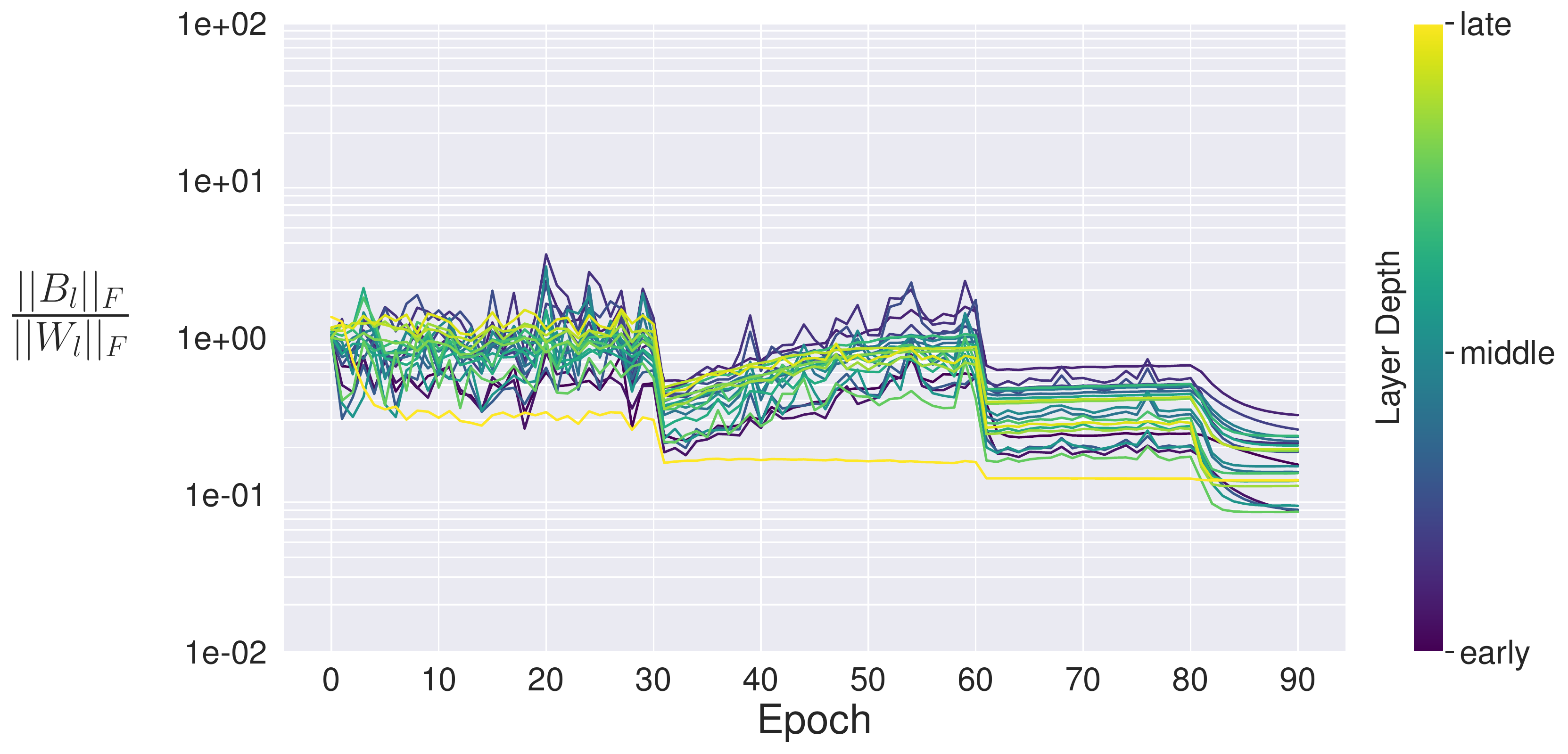}
    \caption{Information Alignment}
\end{subfigure}
\caption{\textbf{Weight metrics during training.} Figures on the left column show the angle between the forward and the backward weights at each layer, depicting their degree of alignment. 
Figures on the right column show the ratio of the Frobenius norm of the backward weights to the forward weights during training. 
For Symmetric Alignment (a) we can clearly see how the weights align very early during training, with the learning rate drops allowing them to further decrease. 
Additionally, the sizes of forward and backwards weight also remain at the same scale during training.
Activation Alignment (b) shows similar behavior to activation, though some of the earlier layers fail to align as closely as the Symmetric Alignment case.
Weight Mirror (c) shows alignment happening within the first few epochs, though some of the later layers don't align as closely.
Looking at the size of the weights during training, we can observe the unstable dynamics explained in \S\ref{sec:local-learning-rules} with exploding and collapsing weight values (Fig.~\ref{fig:stability}) within the first few epochs of training.
Information Alignment (d) shows a similar ordering in alignment as weight mirror, but overall alignment does improve throughout training, with all layers aligning within 5 degrees. 
Compared to weight mirror, the norms of the weights are more stable, with the backward weights becoming smaller than their forward counterparts towards the end of training. 
\label{fig:weight_norms}}
\end{figure*}

In this section we present some visualizations which deepen the understanding of the weight dynamics and stability during training, as presented in \S\ref{sec:local-learning-rules} and \S\ref{sec:non-local-learning-rules}. 
By looking at the weights of the network at each validation point, we are able to compare corresponding forward and backward weights (see Fig.~\ref{fig:full_weight_scatter}) as well as to measure the angle between the vectorized forward and backward weight matrices to quantify their degree of alignment (see Fig.~\ref{fig:weight_norms}). 
Their similarity in terms of scale can also be evaluated by looking at the ratio of the Frobenius norm of the backward weight matrix to the forward weight matrix, $\|B_l\|_F / \|W_l\|_F$.
Separately plotting these metrics in terms of model depth sheds some insight into how different layers behave. 

\section{Further Analysis}
\label{sup:analysis}

\subsection{Instability of Weight Mirror}
\label{sup:stability-analysis}

As explained in \S\ref{sec:local-learning-rules}, the instability of weight mirror can be understood by considering the dynamical system given by the symmetrized gradient flow on $\mathcal{R}_{\text{SA}}$, $\mathcal{R}_{\text{AA}}$, and $\mathcal{R}_{\text{WM}}$ at a given layer $l$.  
By symmetrized gradient flow we imply the gradient dynamics on the loss $\mathcal{R}$ modified such that it is symmetric in both the forward and backward weights.  
We ignore biases and non-linearities and set $\alpha = \beta$ for all three losses.

When the weights, $w_l$ and $b_l$, and input, $x_l$, are all scalar values, the gradient flow for all three losses gives rise to the dynamical system,
$$\frac{\partial}{\partial t}\begin{bmatrix}
w_l\\
b_l
\end{bmatrix} = - A \begin{bmatrix}
w_l\\
b_l
\end{bmatrix},$$
For Symmetric Alignment and Activation Alignment, $A$ is respectively the positive semidefinite matrix
$$A_{\text{SA}} = 
\begin{bmatrix}
1 & -1\\
-1 & 1
\end{bmatrix}
\quad\text{ and }\quad
A_{\text{AA}} = 
\begin{bmatrix}
x_l^2 & -x_l^2\\
-x_l^2 & x_l^2
\end{bmatrix}.$$
For weight mirror, $A$ is the symmetric indefinite matrix
$$
A_{\text{WM}} = 
\begin{bmatrix}
\lambda_{\text{WM}} & -x_l^2\\
-x_l^2 & \lambda_{\text{WM}}
\end{bmatrix}.$$
In all three cases $A$ can be diagonally decomposed by the eigenbasis
$$
\left\{u,v\right\} = \left\{\begin{bmatrix}
1\\
1
\end{bmatrix},
\begin{bmatrix}
1\\
-1
\end{bmatrix}\right\},$$
where $u$ spans the symmetric component and $v$ spans the skew-symmetric component of any realization of the weight vector $\begin{bmatrix}
w_l &
b_l
\end{bmatrix}^\intercal$.

As explained in \S\ref{sec:local-learning-rules}, under this basis, the dynamical system decouples into a system of ODEs governed by the eigenvalues $\lambda_u$ and $\lambda_v$ associated with $u$ and $v$. For all three learning rules, $\lambda_v > 0$ ($\lambda_v$ is respectively $1$, $x^2$, and $\lambda_{\text{WM}} + x_l^2$ for SA, AA, and weight mirror).
For SA and AA, $\lambda_u = 0$, while for weight mirror $\lambda_u = \lambda_{\text{WM}} - x_l^2$.

\subsection{Beyond Feedback Alignment}
\label{sec:analysis-beyond_fa}

An underlying assumption of our work is that certain forms of layer-wise regularization, such as the regularization introduced by Symmetric Alignment, can actually improve the performance of feedback alignment by introducing dynamics on the backward weights. 
To understand these improvements, we build off of prior analyses of backpropagation \cite{saxe_exact_2013} and feedback alignment \cite{baldi_learning_2018}.

Consider the simplest nontrivial architecture: a two layer scalar linear network with forward weights $w_1, w_2$, and backward weight $b$. 
The network is trained with scalar data $\{x_i,y_i\}_{i=1}^n$ on the mean squared error cost function
$$\mathcal{J} = \sum_{i=1}^n\frac{1}{2n}(y_i - w_2w_1x_i)^2.$$
The gradient flow of this network gives the coupled system of differential equations on $(w_1, w_2, b)$
\begin{align}
    \label{dw1}
    \dot{w_1} &= b(\alpha - w_2w_1\beta) \\
    \label{dw2}
    \dot{w_2} &= w_1(\alpha - w_2w_1\beta)
\end{align}
where $\alpha = \sum_{i=1}^n \frac{y_ix_i}{n}$ and $\beta = \sum_{i=1}^n \frac{x_i^2}{n}$. 
For backpropagation the dynamics are constrained to the hyperplane $b = w_2$, while for feedback alignment the dynamics are contained on the hyperplane $b = b(0)$ given by the initialization. 
For Symmetric Alignment, an additional differential equation
\begin{equation}
\label{db}
\dot{b} = w_2 - b,
\end{equation}
attracts all trajectories to the backpropagation hyperplane $b = w_2$. 

To understand the properties of these alignment strategies, we explore the fixed points of their flow. 
From equation (\ref{dw1}) and (\ref{dw2}) we see that both equations are zero on the hyperbola
$$w_2w_1 = \frac{\alpha}{\beta},$$
which is the set of minima of $\mathcal{J}$. 
From equation (\ref{db}) we see that all fixed points of Symmetric Alignment satisfy $b = w_2$. 
Thus, all three alignment strategies have fixed points on the hyperbola of minima intersected with either the hyperplane $b = b(0)$ in the case of feedback alignment or $b = w_2$ in the case of backpropagation and Symmetric Alignment.

In addition to these non-zero fixed points, equation (\ref{dw1}) and (\ref{dw2}) are zero if $b$ and $w_1$ are zero respectively.  
For backpropagation and Symmetric Alignment this also implies $w_2 = 0$, however for feedback alignment $w_2$ is free to be any value. 
Thus, all three alignment strategies have rank-deficient fixed points at the origin $(0,0,0)$ and in the case of feedback alignment more generally on the hyperplane $b = w_1 = 0$.

To understand the stability of these fixed points we consider the local linearization of the vector field by computing the Jacobian matrix\footnote{In the case that the vector field is the negative gradient of a loss, as in backpropagation, then this is the negative Hessian of the loss.}
$$J = \begin{bmatrix}
\partial_{w_1}\dot{w_1} & \partial_{w_1}\dot{w_2} & \partial_{w_1}\dot{b}\\
\partial_{w_2}\dot{w_1} & \partial_{w_2}\dot{w_2} & \partial_{w_2}\dot{b}\\
\partial_{b}\dot{w_1} & \partial_{b}\dot{w_2} & \partial_{b}\dot{b}
\end{bmatrix}.$$
A source of the gradient flow is characterized by non-positive eigenvalues of $J$, a sink by non-negative eigenvalues of $J$, and a saddle by both positive and negative eigenvalues of $J$.

On the hyperbola $w_2w_1 = \frac{\alpha}{\beta}$ the Jacobian matrix for the three alignment strategies have the corresponding eigenvalues:
\begin{table}[H]
\hspace*{-0.25cm}
\centering
\begin{tabular}{@{}lccc@{}} \toprule
    & $\lambda_1$ & $\lambda_2$ & $\lambda_3$ \\ \midrule
    Backprop. & $-\left(w_1^2 + w_2^2\right)x^2$ & $0$ & \\
    Feedback & $-\left(w_1^2 + bw_2\right)x^2$ & $0$ & $0$\\
    Symmetric & $-\left(w_1^2 + w_2^2\right)x^2$ & $0$ & $-1$ \\ \bottomrule
\end{tabular}
\end{table}
Thus, for backpropagation and Symmetric Alignment, all minima of the the cost function $\mathcal{J}$ are sinks, while for feedback alignment the stability of the minima depends on the sign of $w_1^2 + bw_2$. 

From this simple example there are two major takeaways:
\begin{enumerate}
    \item All minima of the cost function $\mathcal{J}$ are sinks of the flow given by backpropagation and Symmetric Alignment, but only some minima are sinks of the flow given by feedback alignment.
    \item Backpropagation and Symmetric Alignment have the exact same critical points, but feedback alignment has a much larger space of rank-deficient critical points.
\end{enumerate}
Thus, even in this simple example it is clear that certain dynamics on the backward weights can have a stabilizing effect on feedback alignment.

\subsection{Kolen-Pollack Learning Rule}
\label{sec:kolen-pollack}

If we consider primitives that are functions of the pseudogradients $\widetilde{\nabla}_{l}$ and $\widetilde{\nabla}_{l+1}$ in addition to the forward weight $W_l$, backward weight $B_l$, layer input $x_l$, and layer output $x_{l+1}$, then the Kolen-Pollack algorithm, originally proposed by \citet{Kolen1994backpropagation} and modified by \citet{akrout_deep_2019}, can be understood in our framework. 

\begin{figure}[t]
\centering
\includegraphics[width=1.0\columnwidth]{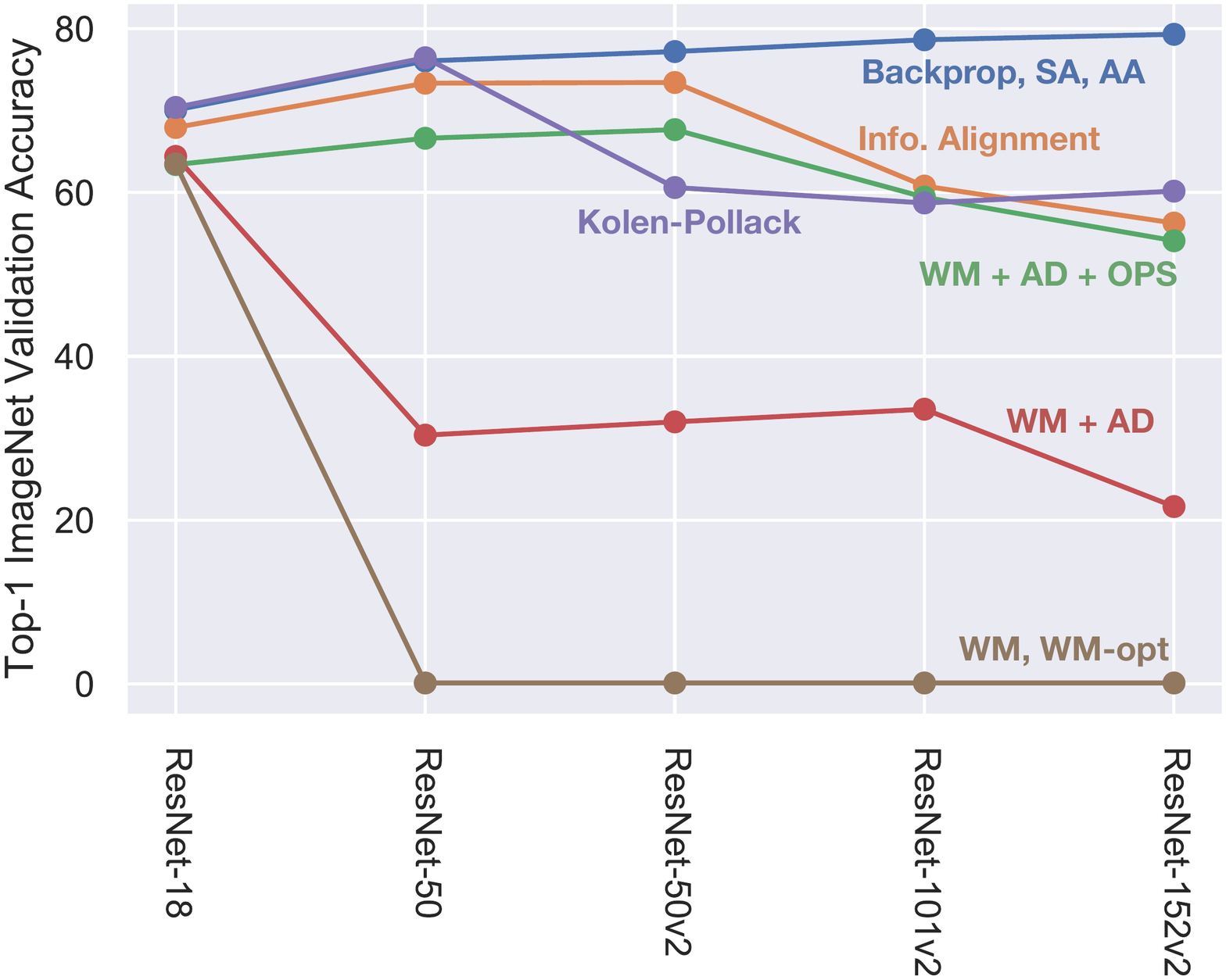}
\caption{\textbf{Performance of Kolen-Pollack across architectures.} We fixed the categorical and continuous metaparameters for ResNet-18 and applied them directly to deeper and different ResNet variants (e.g. v2) as in Fig.~\ref{fig:hp-deeper}. The Kolen-Pollack learning rule, matched backpropagation performance for ResNet-18 and ResNet-50, but a performance gap emerged for different (ResNet-50v2) and deeper (ResNet-101v2, ResNet-152v2) architectures.}
\label{fig:hp-deeper-kp}
\vskip -0.1in
\end{figure}

The Kolen-Pollack algorithm circumvents the weight transport problem, by instead transporting the weight updates and adding weight decay.
Specifically, the forward and backward weights are updated respectively by
\begin{align*}
    \Delta W_l &= -\eta \widetilde{\nabla}_{l+1}x_l^\intercal - \lambda_{\text{KP}} W_l,\\
    \Delta B_l &= -\eta x_l \widetilde{\nabla}_{l+1}^\intercal - \lambda_{\text{KP}} B_l,
\end{align*}
where $\eta$ is the learning rate and $\lambda_{\text{KP}}$ a weight decay constant.
The forward weight update is the standard pseudogradient update with weight decay, while the backward weight update is equivalent to gradient descent on 
$$\mathrm{tr}(x_l^\intercal B_l \widetilde{\nabla}_{l+1}) + \frac{\lambda_{\text{KP}}}{2\eta}||B_l||^2.$$
Thus, if we define the \textit{angle} primitive
$$\mathcal{P}_l^{\text{angle}} = \mathrm{tr}(x_l^\intercal B_l\widetilde{\nabla}_{l+1}) =  \mathrm{tr}(x_l^\intercal \widetilde{\nabla}_{l}),$$
then the \textbf{Kolen-Pollack (KP)} update is given by gradient descent on the layer-wise regularization function
$$\mathcal{R}_{\text{KP}} = \sum_{l \in \text{layers}} \alpha\mathcal{P}^{\text{angle}}_l + \beta\mathcal{P}^{\text{decay}}_l,$$
for $\alpha = 1$ and $\beta = \frac{\lambda_{\text{KP}}}{\eta}$.
The angle primitive encourages alignment of the forward activations with the backward pseudogradients and is local according to the criterion for locality defined in \S\ref{sec:framework-primitives}.
Thus, the Kolen-Pollack learning rule only involves the use of local primitives, but it does necessitate that the backward weight update given by the angle primitive is the exact transpose to the forward weight update at each step of training. 
This constraint is essential to showing theoretically how Kolen-Pollack leads to alignment of the forward and backward weights \cite{Kolen1994backpropagation}, but it is clearly as biologically suspect as exact weight symmetry.
To determine empirically how robust Kolen-Pollack is when loosening this hard constraint, we add random Gaussian noise to each update.
As shown in Fig.~\ref{fig:all-noise}, even with certain levels of noise, the Kolen-Pollack learning rule can still lead to well performing models.
This suggests that a noisy implementation of Kolen-Pollack that removes the constraint of exactness might be biologically feasible.

While Kolen-Pollack uses significantly fewer metaparameters than weight mirror or information alignment, the correct choice of these metaparameters is highly dependent on the architecture. 
As shown in Fig.~\ref{fig:hp-deeper-kp}, the Kolen-Pollack learning rule, with metaparameters specified by \citet{akrout_deep_2019}, matched backpropagation performance for ResNet-18 and ResNet-50.
However, a considerable performance gap with backpropagation as well as our proposed learning rules (information alignment, SA, and AA) emerged for different (ResNet-50v2) and deeper (ResNet-101v2, ResNet-152v2) architectures, providing additional evidence for the necessity of the circuits we propose in maintaining robustness across architecture.

\end{document}